\documentclass[aps,prl,reprint,superscriptaddress,longbibliography,floatfix,footinbib]{revtex4-2}
\usepackage{physics,graphicx,amssymb,bm,microtype,hyperref,siunitx}
\hypersetup{colorlinks=true,linkcolor=black,urlcolor=blue,citecolor=black}
\newcommand{\mr}[1]{\mathrm{#1}}
\newcommand{\ms}[1]{\mathsf{#1}}
\newcommand{\mc}[1]{\mathcal{#1}}
\newcommand{\eq}[1]{\text{$#1$}}

\begin{document}
\title{Inherent electro-optic Kerr rotation}
\author{Erlend Sylju{\aa}sen}
\affiliation{Center for Quantum Spintronics, Department of Physics,
Norwegian University of Science and Technology, NO-7491 Trondheim, Norway}
\author{Alireza Qaiumzadeh}
\affiliation{Center for Quantum Spintronics, Department of Physics,
Norwegian University of Science and Technology, NO-7491 Trondheim, Norway}
\author{Rembert A. Duine}
\affiliation{Institute for Theoretical Physics, Utrecht University, 3584 CC Utrecht, The Netherlands}
\affiliation{Department of Applied Physics, Eindhoven University of Technology, 5600 MB Eindhoven, The Netherlands}
\author{Arne Brataas}
\affiliation{Center for Quantum Spintronics, Department of Physics,
Norwegian University of Science and Technology, NO-7491 Trondheim, Norway}
\begin{abstract}
We uncover a previously overlooked contribution to the electro-optic Kerr rotation of reflected light, arising from the interplay of matter, the static electric field, and the magnetic component of light. This contribution remains nonzero even in isotropic nonmagnetic homogeneous systems. We derive analytical expressions for the Kerr rotation in both two-dimensional layers and semi-infinite systems. Within the relaxation-time approximation, we predict experimentally accessible signal magnitudes in metals. This inherent mechanism thereby opens opportunities for probing electronic properties in materials through Kerr spectroscopy.
\end{abstract}
\maketitle

\textit{Introduction}. Electro- and magneto-optic effects show how static electric and magnetic fields modify the optical properties of materials. Within nonlinear optics, these phenomena appear as birefringence, gyrotropy, dichroism, and related effects. These effects serve as sensitive probes of microscopic interactions in solids and underpin technologies from ultrafast optical modulators to precision field sensors \cite{pershan_1967, haider_2017, wang_2025}. The standard picture attributes them to interactions between matter, the static field, and the electric component of light. Recently, however, unconventional light–matter couplings involving the magnetic component of light have also attracted interest \cite{toupin_1961, kizel_1975, souza_2016, desousa_2024}. Among the many electro- and magneto-optic responses, the Kerr effect stands out: it occurs across a wide range of materials and offers unique sensitivity to subtle symmetry breaking.

The Kerr effect refers here to a rotation in the polarization of light on reflection from a material. Kerr first observed it in ferromagnets \cite{kerr_1877, kerr_1878}, where optical modes of opposite chirality propagate with different refractive indices, leading to a relative phase shift upon reflection \cite{argyres_1955}. The resulting Kerr rotation angle measures the rotation of the reflected polarization axis. Beyond ferromagnets, Kerr signals have exposed exotic phases in superconductors \cite{kapitulnik_2006, kapitulnik_2008, kapitulnik_2018, thomas_2023} and topological insulators \cite{macdonald_2010, tokura_2016}, and more recently enabled detection of the spin and orbital Hall effects (SHE and OHE) \cite{awschalom_2004, lee_2016, gambardella_2017, son_2019, lyalin_2023, choi_2023}. In the SHE (OHE), a static electric field drives a transverse flow of spin (orbital) angular momentum, which accumulates at surfaces as a magnetization detectable through the Kerr effect. This mechanism represents an electro-optic Kerr signal, with rotation angles typically ranging from \eq{10^{-14}} to \eq{10^{-10}} \si{\radian\per\volt\meter} when normalized by the static electric field~\footnote{See table 1 in Ref.~\cite{manchon_2024} for a summary of experimental values.}.

\begin{figure}[htb]
    \centering
    \includegraphics[width=0.49\linewidth]{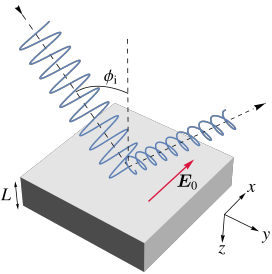}
    \caption{Schematic of electro-optic Kerr rotation. Linearly polarized light with angle of incidence \eq{\phi_\mr{i}} reflects from a material of thickness \eq{L} and becomes elliptically polarized owing to anisotropy induced by an applied static electric field \eq{\bm{E}_0}. The static field may also have a \eq{y}-component.}
    \label{fig:geometry}
\end{figure}

In this Letter, we identify a contribution to the electro-optic Kerr effect that remains nonzero even in isotropic nonmagnetic homogeneous metals. This contribution is therefore inherent to electro-optic Kerr rotation measurements, since it is symmetry-allowed in any material and thus represents a ubiquitous contribution to the measured signal. Microscopically, it stems from an unconventional optical Hall effect driven by the interplay of matter, the static electric field, and the magnetic component of light. In this mechanism, the magnetic field of light itself acts as the time-reversal symmetry breaking perturbation. To expose the physics clearly, we analyze two limits: a two-dimensional (2D) material and a semi-infinite bulk. For a 2D material embedded in a uniform medium, we derive a compact expression for the Kerr rotation angle in the transparent limit,
\begin{equation} \label{eq:theta_s_thin}
    \theta_s \approx  \frac{1}{2} \sin(2\phi_\mr{i}) \frac{v_\mr{d}}{c} ,
\end{equation}
where \eq{\phi_\mr{i}} is the incident angle of light with electric field normal to the plane of incidence (\eq{s}-polarized), and the drift velocity \eq{v_\mr{d}} is linear in the static electric field oriented along the polarization direction (see Fig.~\ref{fig:geometry}). The ratio of the drift velocity to the speed of light in the surrounding medium \eq{c} captures the unconventional response driven by the magnetic component of light. This component is weaker than the electric component by exactly this factor, yet it generates a distinct Kerr signal. In the remainder of this Letter, we derive the inherent contribution to the electro-optic Kerr effect, identify the conditions under which it vanishes, and estimate its magnitude for representative materials.

\textit{Setup}. We consider a static homogeneous electric field \eq{\bm{E}_0} together with a monochromatic electromagnetic plane wave incident on a material. The setup is shown in Fig.~\ref{fig:geometry}: the top surface lies in the \eq{z = 0} plane, \eq{\bm{E}_0} is confined to the \eq{xy}-plane \eq{(E^z_0 = 0)}, the plane of incidence is the \eq{yz}-plane, the angle of incidence is \eq{\phi_\mr{i}}, and the material thickness is \eq{L}. We focus on the optical regime, where the wavelength \eq{\lambda} of the incident wave greatly exceeds the interatomic spacing in the material. To capture the limiting behavior of the Kerr effect, we analyze two cases: a 2D material obtained by taking \eq{L \to 0^+}, and a semi-infinite bulk obtained by fixing the top surface of the material at \eq{z=0} and taking \eq{L \to \infty}. Together, these limits bracket the full range of realistic geometries, from atomically thin layers to extended bulk solids.

The complex Kerr angle \eq{\Theta} characterizes the polarization change of light reflected from a material. For an incident wave with polarization either perpendicular (\textit{senkrecht} or \eq{s}) or parallel (\eq{p}) to the plane of incidence, \eq{\Theta} takes a simple form in the small rotation limit \cite{zak_1990, shin_1996},
\begin{subequations}
\begin{align}
    \Theta_s = \theta_s + i \epsilon_s &\approx \frac{r_{ps}}{r_{ss}},
    \\[0.2cm]
    \Theta_p = \theta_p + i \epsilon_p &\approx \frac{r_{sp}}{r_{pp}},
\end{align}
\end{subequations}
where the complex reflection matrix relates the amplitude of the reflected \eq{(\mr{r})} and incident \eq{(\mr{i})} waves via 
\begin{equation}
    \begin{pmatrix}
        \mc{E}^s_\mr{r} \\ \mc{E}^p_\mr{r}
    \end{pmatrix}
    =
    \begin{pmatrix}
        r_{ss} & r_{sp} 
        \\
        r_{ps} & r_{pp}
    \end{pmatrix}
    \begin{pmatrix}
        \mc{E}^s_\mr{i} \\ \mc{E}^p_\mr{i}
    \end{pmatrix}.
\end{equation}
The Kerr rotation angle \eq{\theta} measures the tilt between the incident polarization axis and the major axis of the reflected polarization ellipse. The ellipticity \eq{\epsilon = \arctan(b / a)} quantifies its shape by giving the ratio of the minor axis \eq{b} to the major axis \eq{a} of the reflected polarization ellipse~\cite{oppeneer_1999}.

To obtain the reflection matrix and hence the complex Kerr angle, we use a macroscopic formulation of Maxwell's equations tailored to optical frequencies \cite{landau_1984}. Outside the material, we model possible surrounding substrates with a linear frequency-independent isotropic relation for the displacement field: \eq{\bm{D}_\pm = \varepsilon_\pm \bm{E}_\pm}. Here, \eq{\bm{E}_\pm} is the electric field and \eq{\varepsilon_\pm} is the dielectric constant in the \eq{z > L} \eq{(+)} or \eq{z < 0} \eq{(-)} region. Inside the material, we express the induced charge current \eq{\bm{J}} through the constitutive relation with the electromagnetic fields,
\begin{align} \label{eq:constitutive_relation}
    &\bm{J}(\bm{r}, t) = \sigma_\mr{E} \ast \bm{E} + \sigma_\mr{H} \ast \bm{H}
    \\
    &+\sigma_\mr{EE} \star  (\bm{E} \otimes \bm{E}) + \sigma_\mr{HH} \star (\bm{H} \otimes \bm{H}) + \sigma_\mr{EH} \star (\bm{E} \otimes \bm{H}) \notag,
\end{align}
with conductivity tensors denoted by \eq{\sigma}. The symbol \eq{\otimes} represents the tensor product, while \eq{*} and \eq{\star} indicate linear and bilinear temporal convolution, respectively, as
\begin{subequations}
\begin{align}
    &(\sigma_\mr{A} \ast \bm{A})^i = \!\! \int \!\! \dd t' \sigma^{ij}_\mr{A}(t - t') A^j(\bm{r}, t'),
    \\
    &\left[ \sigma_\mr{AB} \star (\bm{A} \otimes \bm{B})\right]^i = \int \!\! \dd t' \!\! \int \!\! \dd t'' \sigma^{ijk}_\mr{AB}(t - t', t - t'') \notag \\
    &\hspace{4cm} 
    \times A^j(\bm{r}, t') B^k(\bm{r}, t''), 
\end{align}
\end{subequations}
for two vector fields \eq{\bm{A}} and \eq{\bm{B}} with repeated spatial indices \eq{i, j, k \in \{x,y,z\}} implicitly summed over. The linear electric conductivity \eq{\sigma_\mr{E}} encodes the familiar Ohm’s law, while the linear magneto-conductivity \eq{\sigma_\mr{H}} is far less common \cite{kharzeev_2014, brien_2017} and will vanish in our description. To capture how a static electric field \eq{\bm{E}_0} alters the Kerr effect at leading order, we expand the response to second order in fields. This introduces nonlinear electric contributions from \eq{\sigma_\mr{EE}}, nonlinear magnetic contributions from \eq{\sigma_\mr{HH}} (which also will vanish), and a mixed electro-magneto nonlinear conductivity \eq{\sigma_\mr{EH}}. This constitutive relation thus accounts for unconventional couplings to both the electric and magnetic fields. We neglect equilibrium currents and treat the conductivity tensors as spatially uniform, since the wavelength of the incident wave far exceeds the interatomic scale of the material.

To simplify the analysis, we neglect second and higher-order harmonic terms in the electric and magnetic fields. The charge current inside the material then separates into a static part involving \eq{\bm{E}_0} and a wave part oscillating with the incident wave frequency \eq{\omega}. The wave part includes the influence of \eq{\bm{E}_0} via the second-order responses. We focus on the simplest case of an isotropic material, which sharply reduces the number of independent elements in the conductivity tensors, since the physical properties of the material must remain invariant under all point-group operations \cite{neumann_1873}. Under spatial inversion, both the electric field and charge current change sign, whereas the magnetic field does not. This leaves only two nonzero conductivity tensors \eq{\sigma_\mr{E}} and \eq{\sigma_\mr{EH}}. Rotational invariance further constrains these tensors to \eq{\sigma^{ij}_\mr{E} = \delta^{ij} \sigma_\mr{e}} and \eq{\sigma^{ijk}_\mr{EH} = \varepsilon^{ijk} \sigma_\mr{eh}}, where \eq{\delta} is the Kronecker delta and \eq{\varepsilon} is the Levi-Civita symbol. Finally, we assume the material is nonmagnetic. In the Supplemental Material (SM) \cite{SM}, we show that these assumptions lead to the wave part \eq{(\mr{w})} of the charge current being
\begin{equation} \label{eq:wave_part_current}
\begin{split}
    \bm{J}_\mr{w}(\bm{r}, t) = \mr{Re} \left\{ \left[\sigma_\mr{e}(\omega) \bm{\mc{E}}_\mr{w} + \sigma_\mr{eh}(0, \omega) \bm{E}_0 \cp \bm{\mc{H}}_\mr{w} \right] e^{i(\bm{k} \cdot \bm{r} - \omega t)} \right\},
\end{split}
\end{equation}
where conductivities are understood to be in Fourier space, and \eq{\bm{\mc{E}}_\mr{w}} (\eq{\bm{\mc{H}}_\mr{w}}) is the complex electric (magnetic) field amplitude of an electromagnetic plane wave with wavevector \eq{\bm{k}} and frequency \eq{\omega}. The mixed conductivity \eq{\sigma_\mr{eh}(0,\omega)} describes the response to the static electric field and magnetic component of the wave, with the zero-frequency value defined as the average of the limits approached from positive and negative frequency \cite{SM}. We emphasize that our following derivation of the Kerr angle in the two material geometries, resulting in Eqs.~\eqref{eq:kerr_s_thin} and \eqref{eq:kerr_s_half_infinite}, applies to any choice of microscopic model and to any method used to compute the relevant conductivities.

\textit{Semiclassical transport}. We use the semiclassical Boltzmann transport equation within the relaxation-time approximation to compute the relevant conductivities. The nonequilibrium distribution function satisfies
\begin{equation}
    \pdv{f}{t} - \frac{e}{\hbar} \left(\bm{E} + \bm{v} \cross \bm{B} \right) \vdot \pdv{f}{\bm{k}} = -\frac{(f - f_0)}{\tau},
\end{equation}
where \eq{f_0} is the Fermi-Dirac distribution and \eq{\tau} is the relaxation time. We neglect Berry-curvature effects because they vanish in the inversion- and time-reversal symmetric setting considered here, and assume a single parabolic band with dispersion \eq{\epsilon = \hbar^2 k^2 / (2m)}, where \eq{m} is the effective mass. The group velocity is \eq{\bm{v} = \nabla_{\bm{k}} \epsilon / \hbar}. We solve the Boltzmann equation recursively in powers of the static and optical fields to extract \eq{\sigma_\mr{e}(\omega)} and \eq{\sigma_\mr{eh}(0,\omega)}. The full derivation is presented in the SM~\cite{SM}. We find
\begin{subequations} \label{eq:conductivities}
\begin{align}
    \sigma_\mr{e}(\omega) &= \frac{n e^2 \tau}{m(1 - i\omega \tau)},
    \\
    \sigma_\mr{eh}(0, \omega) &= -\frac{\mu_0 n e^3 \tau^2}{m^2(1 - i\omega \tau)},
\end{align}
\end{subequations}
where \eq{\mu_0} is the vacuum permeability, \eq{n} is the charge carrier density and \eq{e} is the elementary charge. The linear conductivity \eq{\sigma_\mr{e}(\omega)} is the optical Drude conductivity, and we emphasize that \eq{\sigma_\mr{eh}(0, \omega)} characterizes the response due to the magnetic component of light in the presence of a static electric field. This distinguishes the response from the classical Hall effect that would be described by \eq{\sigma_\mr{eh}(\omega, 0)} to leading order in a static magnetic field. We now proceed to compute the Kerr effect for the two limiting cases of a 2D material and a semi-infinite bulk.

\textit{2D material}. In this limit, the reflection matrix is obtained by matching solutions of Maxwell’s macroscopic equations across linear isotropic frequency-independent media using appropriate boundary conditions at \eq{z = 0}. The material’s electromagnetic response enters through the boundary conditions governing the in-plane components of the magnetic field,
\begin{equation} \label{eq:thin_boundary_condition}
    \hat{\bm{z}} \cp \left[ \bm{H}(\bm{r}, t)|_{z = 0^+} - \bm{H}(\bm{r}, t)|_{z = 0^-} \right] = \bm{K}(\bm{r}, t),
\end{equation}
where the surface current \eq{\bm{K}} appears as a delta function contribution to the current, \eq{\bm{J} = \delta(z) \bm{K}}. The wave part of the surface current is related to the electric and magnetic fields via Eq.~\eqref{eq:wave_part_current}, where we in the constitutive relation use the averaged fields \eq{\bm{E} \to \bar{\bm{E}} = (\bm{E}|_{z=0^+} + \bm{E}|_{z=0^-}) / 2}, and similarly \eq{\bm{H} \to \bar{\bm{H}}}. The magnetic field is discontinuous across \eq{z=0} as determined by Eq.~\eqref{eq:thin_boundary_condition}, so a prescription for the field at the interface is mandatory. We use the averaged fields, which do not favor either side of the interface. This also conveniently cancels the contribution from the static magnetic field generated by the induced static current due to \eq{\bm{E}_0}, as these static magnetic fields point in opposite directions above and below the 2D material. The full computation of the reflection matrix and complex Kerr angle for the 2D material is performed in the SM \cite{SM}. 

\begin{figure}[h]
    \includegraphics[width = \linewidth]{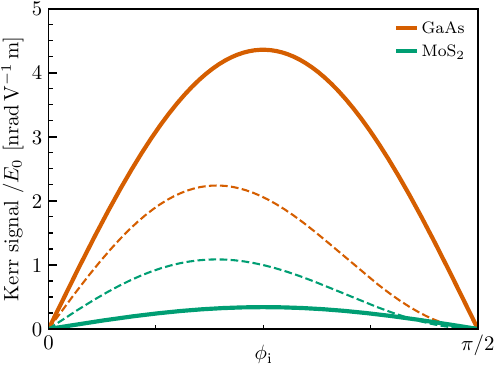}
	\caption{Kerr rotation \eq{\theta_s / E_0} (solid lines) and negative ellipticity \eq{-\epsilon_s / E_0} (dashed lines) vs. angle of incidence for \eq{\mr{GaAs}} and \eq{\mr{MoS_2}}. The wavelength of incident light is set to \eq{\lambda = 800 \; \si{\nano\meter}}, the static electric field is \eq{\bm{E}_0 = \hat{\bm{x}} E_0}, and the surrounding medium has dielectric constant \eq{\varepsilon_\pm = \varepsilon_0}, where \eq{\varepsilon_0} is the vacuum permittivity. The material parameters used are detailed in the SM \cite{SM}. To display all curves on the same axis and highlight the angular dependence, the following scale factors are applied: \eq{\epsilon^\mr{GaAs}_s \times 10^4} and \eq{\epsilon^\mr{MoS_2}_s \times 10^5}. The actual physical values are smaller than the plotted curves by these factors.}
    \label{fig:kerr_signal_thin}
\end{figure}

For \eq{s}-polarized incident light with \eq{\bm{E}_0 = \hat{\bm{x}} E_0}, we find the complex Kerr angle
\begin{equation} \label{eq:kerr_s_thin}
	\Theta_s = \frac{2 \sin(\phi_\mr{i}) \sigma_\mr{eh}(0, \omega) E_0}{f_1 f_2},
\end{equation}
where
\begin{subequations}
\begin{align}
	f_1 &= 1 + \frac{n_+ \cos(\phi_\mr{i})}{n_- \cos(\phi_\mr{t})} + \sigma_\mr{e}(\omega) Z_- \cos(\phi_\mr{i}),
	\\
	f_2 &= 1 - \frac{n_+ \cos(\phi_\mr{t}) }{n_- \cos(\phi_\mr{i})} - \sigma_\mr{e}(\omega) Z_- \sec(\phi_\mr{i}).
\end{align}
\end{subequations}
The transmitted angle \eq{\phi_\mr{t}} is related to the incident angle by Snell's law: \eq{ n_+ \sin(\phi_\mr{t}) = n_- \sin(\phi_\mr{i})}, where the refractive index for \eq{\pm z > 0} is \eq{n_\pm = \sqrt{\varepsilon_\pm}} and the wave impedance is \eq{Z_\pm = \sqrt{\mu_0 / \varepsilon_\pm}}. Equation~\eqref{eq:kerr_s_thin} shows that the complex Kerr angle is linear in both the static electric field and the unconventional conductivity \eq{\sigma_\mr{eh}(0, \omega)}. Figure~\ref{fig:kerr_signal_thin} presents the Kerr rotation and ellipticity angles estimated for two representative 2D systems: a gallium arsenide-based two-dimensional electron system \eq{(\mr{GaAs})}, and monolayer molybdenum disulfide \eq{(\mr{MoS_2})}. 

To gain analytical insight, we consider identical media on both sides of the material, reflected by \eq{\varepsilon_\pm = \varepsilon}. We further assume \eq{\sigma'_\mr{e}(\omega) Z \ll 2} and \eq{\sigma''_\mr{e}(\omega) Z \ll 2}, where the linear conductivity is decomposed into real and imaginary parts as \eq{\sigma_\mr{e}(\omega) = \sigma'_\mr{e} + i \sigma''_\mr{e}} with \eq{\sigma'_\mr{e}, \sigma''_\mr{e} \in \mathbb{R}}. The product \eq{\sigma_\mr{e} Z} is dimensionless in 2D and compares the material's linear electric conductivity to the resistance of electromagnetic wave propagation in the surrounding medium, which is for the 2D materials and the frequency considered much less than unity \cite{SM}. This regime also corresponds to the transparent limit \cite{li_2018, SM}, in which inserting the conductivities in Eq.~\eqref{eq:conductivities} into Eq.~\eqref{eq:kerr_s_thin} gives Eq.~\eqref{eq:theta_s_thin}, where \eq{v_\mr{d} = \mu E_0} is the drift velocity with carrier mobility \eq{\mu = e \tau / m} and \eq{c = 1 / \sqrt{\mu_0 \varepsilon}} is the speed of light in the surrounding medium. Equation~\eqref{eq:theta_s_thin} explains why \eq{\mr{GaAs}} exhibits stronger Kerr rotation than \eq{\mr{MoS_2}} in Fig.~\ref{fig:kerr_signal_thin}, as its electrons have larger mobility and therefore higher drift velocity \cite{SM}.

As shown in the SM \cite{SM}, \eq{p}-polarized incident light produces no Kerr signal \eq{(\Theta_p = 0)}, independent of the static electric field orientation and angle of incidence. The reason is that both the induced current and the static electric field lie in the \eq{xy}-plane. The antisymmetric form of the second-order response tensor \eq{\sigma_\mr{EH}} restricts the system’s response to the \eq{z}-component of the magnetic field of the wave, which is finite only for \eq{s}-polarized incident light at oblique incidence. Furthermore, we find that the combination of \eq{s}-polarized incident light and static electric field oriented in the plane of incidence gives no Kerr signal \cite{SM}. In this configuration, the unconventional current in Eq.~\eqref{eq:wave_part_current} is parallel with the linear current, thereby only renormalizing the longitudinal response without the polarization mixing required for a finite Kerr signal. For a general static electric field \eq{\bm{E}_0 = \hat{\bm{x}} E^x_0 + \hat{\bm{y}} E^y_0}, we find for the same reason that the \eq{y}-component contributes only as a small correction to the expression in Eq.~\eqref{eq:theta_s_thin} \cite{SM}. We also examine the influence of asymmetric media above and below the material \eq{(\varepsilon_+ \neq \varepsilon_-)}, which is found to suppress the Kerr signal significantly to rotation angles per unit static electric field in the order of \eq{10^{-16}} to \eq{10^{-11}} \si{\radian\per\volt\meter} \cite{SM}. These results identify suspended 2D materials as the most promising platforms for observing this effect.

\textit{Semi-infinite bulk}. We now turn to the opposite limit, where the material thickness far exceeds the incident wavelength. In this case, we fix the upper surface of the material at \eq{z = 0} and send \eq{L \to \infty} in Fig.~\ref{fig:geometry}, so internal reflections can be neglected. Inside the material, the plane-wave solution to Maxwell’s macroscopic equations follows from the dispersion relation for the normal component of the wavevector, since the in-plane components must match those outside to satisfy the boundary conditions at the interface. We obtain \eq{k_z = C \omega / v_+} \cite{SM}, where \eq{k_z} is the normal component of the wavevector inside the material, \eq{v_+ = 1 / \sqrt{\mu_0 \varepsilon_\mr{e}(\omega)}} is the speed of light inside the material, and the anisotropic correction is
\begin{equation}
	C^2 = 1 - \sin^2(\phi_\mr{i}) \frac{\varepsilon_-}{\varepsilon_\mr{e}(\omega)} - \sin(\phi_\mr{i}) \frac{\varepsilon_\mr{eh}(0, \omega) E^y_0}{\varepsilon_\mr{e}(\omega) Z_-}.
\end{equation}
The permittivities are defined as \eq{\varepsilon_\mr{e}(\omega) = \varepsilon_0 + i \sigma_\mr{e}(\omega) / \omega} and \eq{\varepsilon_\mr{eh}(0, \omega) = i \sigma_\mr{eh}(0, \omega) / \omega}. The reflection matrix is then obtained from Maxwell’s boundary conditions \cite{SM}, where we assume no special surface effects to represent a bulk material.

\begin{figure}[h]
	\includegraphics[width = \linewidth]{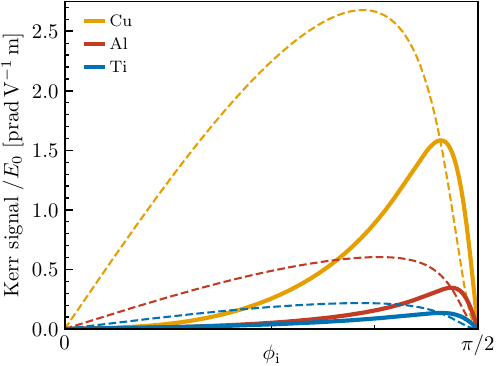}
	\caption{Kerr rotation \eq{\theta_s / E_0} (solid lines) and negative ellipticity \eq{-\epsilon_s / E_0} (dashed lines) vs. angle of incidence for copper \eq{(\mr{Cu})}, titanium \eq{\mr{(Ti)}}, and aluminum \eq{(\mr{Al})}, modeled as semi-infinite bulk materials. The wavelength of incident light is set to \eq{\lambda = 800 \;\si{nm}}, the static electric field is \eq{\bm{E}_0 = \hat{\bm{x}} E_0}, and the dielectric constant above the material is \eq{\varepsilon_- = \varepsilon_0}. The material parameters used are detailed in the SM \cite{SM}.}
    \label{fig:kerr_signal_half_infinite}
\end{figure}

Similarly as for the 2D material, we find \eq{\Theta_p = 0} for any \eq{\phi_\mr{i}} and any orientation of \eq{\bm{E}_0}, \eq{\Theta_s = 0} when \eq{\bm{E}_0 = \hat{\bm{y}} E_0}, and \eq{\Theta_s(E^x_0, E^y_0) \approx \Theta_s(E^x_0, 0)}. For \eq{\bm{E}_0 = \hat{\bm{x}} E_0}, we obtain
\begin{equation} \label{eq:kerr_s_half_infinite}
    \Theta_s = \frac{2v_+ \sin(\phi_\mr{i}) \varepsilon_\mr{eh}(0, \omega) E_0}{C g_1 g_2}, 
\end{equation}
where \eq{C} is here evaluated at \eq{E^y_0 = 0},
\begin{subequations}
\begin{align}
    g_1 &= 1 - \frac{n_+ C}{n_- \cos(\phi_\mr{i})},
    \\
    g_2 &= 1 + \frac{n_+ \cos(\phi_\mr{i})}{n_- C},
\end{align}
\end{subequations}
and the refractive index inside the material is \eq{n_+ = \sqrt{\varepsilon_\mr{e}(\omega)}}. The magnitude of the Kerr rotation angle and ellipticity is estimated for three metals in Fig.~\ref{fig:kerr_signal_half_infinite}. The overall magnitudes are smaller than for the 2D materials but remain experimentally accessible, with their maxima shifted to larger angles of incidence. We show in the SM \cite{SM} that varying the dielectric constant above the semi-infinite bulk gives only small modifications to the magnitudes. 

\textit{Concluding remarks}. We have revealed a contribution to the electro-optic Kerr rotation that stems from the coupling between matter, the static electric field and the magnetic component of light. This is an inherent mechanism because it remains nonzero even in isotropic nonmagnetic homogeneous metal, and must therefore be considered when interpreting Kerr measurements, for example in experiments probing spin and orbital Hall effects. Our analysis of 2D and semi-infinite metals shows that the effect appears only at oblique incidence, with the static electric field and incident polarization oriented perpendicular to the plane of incidence. The predicted Kerr signal magnitudes for representative 2D and bulk metals are well within experimental reach. Two-dimensional systems with large drift velocity, such as GaAs-based two-dimensional systems offer particularly strong Kerr rotation, with predicted angles normalized by the static electric field on the order of \si{\nano \radian\per\volt\meter}. In contrast, bulk metals like aluminum show predicted Kerr rotation on the order of \si{\pico\radian\per\volt\meter}, and display weak spin and orbital Hall signals \cite{salemi_2022, go_2024}, thereby providing clean platforms to disentangle this mechanism from other contributions to the Kerr rotation. 
We used the Boltzmann transport equation within the relaxation-time approximation to obtain these estimates, where the response we consider is classical (kinetic) in origin and arises from field-induced changes in the nonequilibrium distribution function. For systems where quantum-geometric contributions are symmetry allowed, the total response also includes Berry-curvature and orbital-moment contributions as discussed in Ref.~\cite{desousa_2024}. Future work should explore higher-order harmonics, internal reflections in the bulk limit, and refined models of the unconventional Hall conductivity to clarify how band structure, correlations, and disorder shape the strength of this effect.

\textit{Note added}. After our initial submission, Ref.~\cite{mahfouzi_2025} also identified the contribution discussed here as part of their extrinsic bulk electro-optical response.

\section*{Acknowledgments}
A.Q. thanks Mathias Kläui for useful discussions. We acknowledge support from the Research Council of Norway through Grant No. 353919 ``QTransMag'' and Grant No. 262633 ``Center of Excellence on Quantum \nobreak Spintronics''.

\bibliography{references}
\onecolumngrid
\begin{center}
  \textbf{\large Supplemental Material for ``Inherent electro-optic Kerr rotation''}
\end{center}
\setcounter{equation}{0}
\renewcommand{\theequation}{S\arabic{equation}}
\section{Second-order response} \label{appendix:second_order_response}
In this section, we express the second-order response of a real-valued vector field \eq{\bm{F}} to a perturbation composed of a static homogeneous electric field, a static homogeneous magnetic field, and a monochromatic electromagnetic plane wave. We use four-vector notation, where the four-position is \eq{\ms{x} = (\bm{r}, t)} with spatial position \eq{\bm{r}} and time coordinate \eq{t}, and the four-wavevector is \eq{\ms{k} = (\bm{k}, \omega)} with wavevector \eq{\bm{k}} and angular frequency \eq{\omega}. Their scalar product is defined as \eq{\ms{k} \vdot \ms{x} = \bm{k} \vdot \bm{r} - \omega t}. We denote spatial components by superscripts with \eq{i, j, k \in \{x, y, z\}}. Repeated indices within a single term are understood to be summed over all three spatial components. The second-order response is written as
\begin{align} \label{eqa:second_order_response}
    &F^i(\ms{x}) = F^i_{0}(\ms{x}) + \!\! \int \!\! \dd \ms{x}' \left[ \chi_\mr{E}^{ij}(\ms{x} - \ms{x}') E^j(\ms{x}') + \chi_\mr{H}^{ij}(\ms{x} - \ms{x}') H^j(\ms{x}') \right] +
    \\
    &\!\! \int \!\! \dd \ms{x}' \!\! \int \!\! \dd \ms{x}'' \left[ \chi_{\mr{EE}}^{ijk}(\ms{x} - \ms{x}', \ms{x} - \ms{x}'') E^j(\ms{x}') E^k(\ms{x}'') + \chi_{\mr{HH}}^{ijk}(\ms{x} - \ms{x}', \ms{x} - \ms{x}'') H^j(\ms{x}') H^k(\ms{x}'') + \chi^{ijk}_{\mr{EH}}(\ms{x} - \ms{x}', \ms{x} - \ms{x}'') E^j(\ms{x}') H^k(\ms{x}'') \right], \notag
\end{align}
where \eq{\bm{F}_{\mr{0}}} is the equilibrium value and \eq{\chi} are susceptibility tensors. We consider an electromagnetic field consisting of a low-frequency part with four-wavevector \eq{\ms{k}_\mr{s}} and a wave part with four-wavevector \eq{\ms{k}_\mr{w}}, 
\begin{align} 
    \bm{E}(\ms{x}) &= \Re \left( \bm{\mc{E}}_\mr{s} e^{i\ms{k}_\mr{s} \vdot \ms{x}} + \bm{\mc{E}}_\mr{w} e^{i\ms{k}_\mr{w} \vdot \ms{x}} \right), \label{eqa:electromagnetic_fields1}
    \\
    \bm{H}(\ms{x}) &= \Re \left( \bm{\mc{H}}_\mr{s} e^{i\ms{k}_\mr{s} \vdot \ms{x}} + \bm{\mc{H}}_\mr{w} e^{i\ms{k}_\mr{w} \vdot \ms{x}} \right).
    \label{eqa:electromagnetic_fields2}
\end{align}
At the end of the calculations, we will let the low-frequency contribution become static and homogeneous. We define the Fourier transform, and its inverse, of a function \eq{f} as
\begin{align}   
    f(\ms{x}) &= \!\! \int \!\! \frac{\dd \ms{k}}{(2\pi)^4} \,  e^{i \ms{k} \vdot \ms{x}} f(\ms{k}), 
    \\
    f(\ms{k}) &= \!\! \int \!\! \dd \ms{x} \, e^{-i\ms{k} \vdot \ms{x}} f(\ms{x}).
\end{align}
Inserting Eqs.~(\ref{eqa:electromagnetic_fields1}, \ref{eqa:electromagnetic_fields2}) into Eq.~\eqref{eqa:second_order_response} and using the inverse Fourier transform yields
\begin{align}
    F^i(\ms{x}) &= F^i_0(\ms{x}) + 
    \Re \Big\{ 
    \left[ \chi^{ij}_\mr{E}(\ms{k}_\mr{s}) \mc{E}^j_\mr{s} + \chi^{ij}_\mr{H}(\ms{k}_\mr{s}) \mc{H}^j_\mr{s}  \right] e^{i\ms{k}_\mr{s} \vdot \ms{x}} 
    +
    \left[ \chi^{ij}_\mr{E}(\ms{k}_\mr{w}) \mc{E}^j_\mr{w} + \chi^{ij}_\mr{H}(\ms{k}_\mr{w}) \mc{H}^j_\mr{w}  \right] e^{i\ms{k}_\mr{w} \vdot \ms{x}} 
    \\
    &+ \frac{1}{2} \Big[ \chi^{ijk}_\mr{EE}(\ms{k}_\mr{s}, -\ms{k}_\mr{s}) \mc{E}^j_\mr{s} \mc{E}^{k*}_\mr{s} +  \chi^{ijk}_\mr{HH}(\ms{k}_\mr{s}, -\ms{k}_\mr{s}) \mc{H}^j_\mr{s} \mc{H}^{k*}_\mr{s} + \chi^{ijk}_\mr{EH}(\ms{k}_\mr{s}, -\ms{k}_\mr{s}) \mc{E}^j_\mr{s} \mc{H}^{k*}_\mr{s} \notag
    \\
    &\hspace{0.4cm} + \chi^{ijk}_\mr{EE}(\ms{k}_\mr{w}, -\ms{k}_\mr{w}) \mc{E}^j_\mr{w} \mc{E}^{k*}_\mr{w} + \chi^{ijk}_\mr{HH}(\ms{k}_\mr{w}, -\ms{k}_\mr{w}) \mc{H}^j_\mr{w} \mc{H}^{k*}_\mr{w} + \chi^{ijk}_\mr{EH}(\ms{k}_\mr{w}, -\ms{k}_\mr{w}) \mc{E}^j_\mr{w} \mc{H}^{k*}_\mr{w} \Big] \notag
    \\
    &+ \frac{1}{2} \left[ \chi^{i(jk)}_\mr{EE}(\ms{k}_\mr{s}, \mr{k}_\mr{w}) \mc{E}^j_\mr{s} \mc{E}^k_\mr{w} + \chi^{i(jk)}_\mr{HH}(\ms{k}_\mr{s}, \mr{k}_\mr{w}) \mc{H}^j_\mr{s} \mc{H}^k_\mr{w} + \chi^{ijk}_\mr{EH}(\ms{k}_\mr{s}, \mr{k}_\mr{w})\mc{E}^j_\mr{s} \mc{H}^k_\mr{w} + \chi^{ijk}_\mr{EH}(\ms{k}_\mr{w}, \mr{k}_\mr{s})\mc{E}^j_\mr{w} \mc{H}^k_\mr{s} \right] e^{i(\ms{k}_\mr{w} + \ms{k}_\mr{s}) \vdot \ms{x}} \notag 
    \\
    &+ \frac{1}{2} \left[ \chi^{i(jk)}_\mr{EE}(\ms{k}_\mr{w}, -\mr{k}_\mr{s}) \mc{E}^j_\mr{w} \mc{E}^{k*}_\mr{s} + \chi^{i(jk)}_\mr{HH}(\ms{k}_\mr{w}, -\mr{k}_\mr{s}) \mc{H}^j_\mr{w} \mc{H}^{k*}_\mr{s} + \chi^{ijk}_\mr{EH}(-\ms{k}_\mr{s}, \mr{k}_\mr{w})\mc{E}^{j*}_\mr{s} \mc{H}^k_\mr{w} + \chi^{ijk}_\mr{EH}(\ms{k}_\mr{w}, -\mr{k}_\mr{s})\mc{E}^j_\mr{w} \mc{H}^{k*}_\mr{s} \right] e^{i(\ms{k}_\mr{w} - \ms{k}_\mr{s}) \vdot \ms{x}} \notag
    \\
    &+ \frac{1}{2} \left[ \chi^{ijk}_\mr{EE}(\ms{k}_\mr{s}, \ms{k}_\mr{s}) \mc{E}^j_\mr{s} \mc{E}^k_\mr{s} + \chi^{ijk}_\mr{HH}(\ms{k}_\mr{s}, \ms{k}_\mr{s}) \mc{H}^j_\mr{s} \mc{H}^k_\mr{s} + \chi^{ijk}_\mr{EH}(\ms{k}_\mr{s}, \ms{k}_\mr{s})\mc{E}^j_\mr{s} \mc{H}^k_\mr{s} \right] e^{2i\ms{k}_\mr{s} \vdot \ms{x}} \notag
    \\
    &+ \frac{1}{2} \left[ \chi^{ijk}_\mr{EE}(\ms{k}_\mr{w}, \ms{k}_\mr{w}) \mc{E}^j_\mr{w} \mc{E}^k_\mr{w} + \chi^{ijk}_\mr{HH}(\ms{k}_\mr{w}, \ms{k}_\mr{w}) \mc{H}^j_\mr{w} \mc{H}^k_\mr{w} + \chi^{ijk}_\mr{EH}(\ms{k}_\mr{w}, \ms{k}_\mr{w})\mc{E}^j_\mr{w} \mc{H}^k_\mr{w} \right] e^{2i\ms{k}_\mr{w} \vdot \ms{x}}  \Big\}, \notag
\end{align}
where \eq{\chi^{i(jk)}(\ms{k}_1, \ms{k}_2) = \chi^{ijk}(\ms{k}_1, \ms{k}_2) + \chi^{ikj}(\ms{k}_2, \ms{k}_1)}. We next perform the static and homogeneous limit of the low-frequency contribution, where we first take the homogeneous limit \eq{\bm{k}_\mr{s} \to \bm{0}^+}, and then the static limit \eq{\omega_\mr{s} \to 0^+}, which together are written compactly as \eq{\ms{k}_\mr{s} \to 0^+}. The result is
\begin{equation}
    \lim_{\ms{k}_\mr{s} \to 0^+} \bm{F}(\ms{x}) = \bm{F}_0(\ms{x}) + \Re \big( \bm{F}_\mr{s} + \bm{F}_\mr{w} e^{i \ms{k}_\mr{w} \vdot \ms{x}} + \bm{F}_\mr{ww} e^{2i\ms{k}_\mr{w} \vdot \ms{x}} \big),
\end{equation}
where
\begin{align}
    F^i_\mr{s} &= \chi^{ij}_\mr{E}(0^+) \mc{E}^j_\mr{s} + \chi^{ij}_\mr{H}(0^+) \mc{H}^j_\mr{s} 
    \\
    &+ \frac{1}{2} \Big[ \chi^{ijk}_\mr{EE}(0^+, 0^-) \mc{E}^j_\mr{s} \mc{E}^{k*}_\mr{s} + \chi^{ijk}_\mr{EE}(\ms{k}_\mr{w}, -\ms{k}_\mr{w}) \mc{E}^j_\mr{w} \mc{E}^{k*}_\mr{w} + \chi^{ijk}_\mr{EE}(0^+, 0^+) \mc{E}^j_\mr{s} \mc{E}^{k}_\mr{s} \notag
    \\
    &\hspace{0.4cm} + \chi^{ijk}_\mr{HH}(0^+, 0^-) \mc{H}^j_\mr{s} \mc{H}^{k*}_\mr{s} + \chi^{ijk}_\mr{HH}(\ms{k}_\mr{w}, -\ms{k}_\mr{w}) \mc{H}^j_\mr{w} \mc{H}^{k*}_\mr{w}  + \chi^{ijk}_\mr{HH}(0^+, 0^+) \mc{H}^j_\mr{s} \mc{H}^k_\mr{s} \notag
    \\
    &\hspace{0.4cm} + \chi^{ijk}_\mr{EH}(0^+, 0^-) \mc{E}^j_\mr{s} \mc{H}^{k*}_\mr{s} + \chi^{ijk}_\mr{EH}(\ms{k}_\mr{w}, -\ms{k}_\mr{w}) \mc{E}^j_\mr{w} \mc{H}^{k*}_\mr{w} + \chi^{ijk}_\mr{EH}(0^+, 0^+)\mc{E}^j_\mr{s} \mc{H}^k_\mr{s}  \Big], \notag
    \\[0.1cm]
    F^i_\mr{w} &= \chi^{ij}_\mr{E}(\ms{k}_\mr{w}) \mc{E}^j_\mr{w} + \chi^{ij}_\mr{H}(\ms{k}_\mr{w}) \mc{H}^j_\mr{w} 
    \\
    &+ \frac{1}{2} \Big[ \chi^{i(jk)}_\mr{EE}(0^+, \ms{k}_\mr{w}) \mc{E}^j_\mr{s} \mc{E}^k_\mr{w} + \chi^{i(jk)}_\mr{EE}(0^-, \ms{k}_\mr{w}) \mc{E}^{j*}_\mr{s} \mc{E}^k_\mr{w} + \chi^{ijk}_\mr{EH}(\ms{k}_\mr{w}, 0^+) \mc{E}^j_\mr{w} \mc{H}^k_\mr{s}  + \chi^{ijk}_\mr{EH}(\ms{k}_\mr{w}, 0^-) \mc{E}^j_\mr{w} \mc{H}^{k*}_\mr{s}  \notag
    \\
    &\hspace{0.4cm} + \chi^{i(jk)}_\mr{HH}(0^+, \ms{k}_\mr{w}) \mc{H}^j_\mr{s} \mc{H}^k_\mr{w} + \chi^{i(jk)}_\mr{HH}(0^-, \ms{k}_\mr{w}) \mc{H}^{j*}_\mr{s} \mc{H}^k_\mr{w} + \chi^{ijk}_\mr{EH}(0^+, \ms{k}_\mr{w}) \mc{E}^j_\mr{s} \mc{H}^k_\mr{w} + \chi^{ijk}_\mr{EH}(0^-, \ms{k}_\mr{w}) \mc{E}^{j*}_\mr{s} \mc{H}^k_\mr{w} \Big], \notag
    \\[0.1cm]
    F^i_\mr{ww} &= \frac{1}{2} \left[ \chi^{ijk}_\mr{EE}(\ms{k}_\mr{w}, \ms{k}_\mr{w}) \mc{E}^j_\mr{w} \mc{E}^k_\mr{w} + \chi^{ijk}_\mr{HH}(\ms{k}_\mr{w}, \ms{k}_\mr{w}) \mc{H}^j_\mr{w} \mc{H}^k_\mr{w} + \chi^{ijk}_\mr{EH}(\ms{k}_\mr{w}, \ms{k}_\mr{w})\mc{E}^j_\mr{w} \mc{H}^k_\mr{w} \right].
\end{align}
The part that is proportional to \eq{\exp(i \ms{k}_\mr{w} \vdot \ms{x})} can be written as
\begin{equation} \label{eqa:renormalized_response}
    F^i_\mr{w} = \chi^{ik}_\mr{wE}(\ms{k}_\mr{w}; \bm{\mc{E}}_\mr{s}, \bm{\mc{H}}_\mr{s}) \mc{E}^k_\mr{w} + \chi^{ik}_\mr{wH}(\ms{k}_\mr{w}; \bm{\mc{E}}_\mr{s}, \bm{\mc{H}}_\mr{s}) \mc{H}^k_\mr{w},
\end{equation}
where the renormalized conductivities are
\begin{align}
	\chi^{ik}_\mr{wE}(\ms{k}_\mr{w}; \bm{\mc{E}}_\mr{s}, \bm{\mc{H}}_\mr{s}) &= \chi^{ik}_\mr{E}(\ms{k}_\mr{w}) + \frac{1}{2} \left[ \chi^{i(jk)}_\mr{EE} (0^+, \ms{k}_\mr{w}) \mc{E}^j_\mr{s} + \chi^{i(jk)}_\mr{EE} (0^-, \ms{k}_\mr{w}) \mc{E}^{j*}_\mr{s} + \chi^{ikj}_\mr{EH}(\mr{k}_\mr{w}, 0^+) \mc{H}^j_\mr{s} + \chi^{ikj}_\mr{EH}(\mr{k}_\mr{w}, 0^-) \mc{H}^{j*}_\mr{s} \right], 
	\\
	\chi^{ik}_\mr{wH}(\ms{k}_\mr{w}; \bm{\mc{E}}_\mr{s}, \bm{\mc{H}}_\mr{s}) &= \chi^{ik}_\mr{H}(\ms{k}_\mr{w}) + \frac{1}{2} \left[ \chi^{i(jk)}_\mr{HH} (0^+, \ms{k}_\mr{w}) \mc{H}^j_\mr{s} + \chi^{i(jk)}_\mr{HH} (0^-, \ms{k}_\mr{w}) \mc{H}^{j*}_\mr{s} + \chi^{ijk}_\mr{EH}(0^+, \mr{k}_\mr{w}) \mc{E}^j_\mr{s} + \chi^{ijk}_\mr{EH}(0^-, \mr{k}_\mr{w}) \mc{E}^{j*}_\mr{s} \right]. 
\end{align}
Equation~(6) of the main text follows from Eq.~\eqref{eqa:renormalized_response} by applying the symmetry restrictions for an isotropic material presented in the main text, while setting \eq{\bm{\mc{H}}_\mr{s} = \bm{0}} with \eq{\bm{\mc{E}}_\mr{s} \in \mathbb{R}^3}, and taking the spatially uniform limit, i.e., evaluating the susceptibilities at \eq{\bm{k}_\mr{w} = \bm{0}}.

\section{Conductivities from Boltzmann transport equation} \label{appendix:boltzmann_equation}
In this section, we compute the nonzero optical conductivities of an isotropic material to first order in static homogeneous electric and magnetic fields using the Boltzmann transport equation. The main text showed that the only nonzero conductivity tensors are \eq{\sigma_\mr{E}} and \eq{\sigma_\mr{EH}}. In addition to the unconventional conductivity \eq{\sigma_\mr{EH}(0, \omega)}, which is relevant for static electric fields, we also compute the classical Hall effect conductivity \eq{\sigma_\mr{EH}(\omega, 0)} to clarify their differences. The Boltzmann transport equation is
\begin{equation} \label{eqb:boltzmann_equation}
    \pdv{f}{t}  + \dot{\bm{r}} \vdot \pdv{f}{\bm{r}}  + \dot{\bm{k}} \vdot \pdv{f}{\bm{k}} = \left( \pdv{f}{t} \right)_{\mr{coll.}},
\end{equation}
where \eq{f = f(\bm{r}, \bm{k}, t)} is the number of particles in the \eq{d}-dimensional phase space volume \eq{\dd^d r \, \dd^d k / (2\pi)^d} around position \eq{\bm{r}} and momentum \eq{\bm{k}} at time \eq{t}. We assume a spatially homogeneous distribution, so \eq{\partial_{\bm{r}} f = \bm{0}}, and use the relaxation time approximation \eq{(\partial_t f)_\mr{coll.} = -(f - f_0) / \tau}, where \eq{\tau} is the relaxation time and \eq{f_0} is the Fermi-Dirac equilibrium distribution function. The semiclassical equation of motion is \eq{\hbar \dot{\bm{k}} = -e(\bm{E} + \bm{v} \cp \bm{B})}, where \eq{e} is the elementary charge, \eq{\bm{E}} is the electric field, \eq{\bm{B}} is the magnetic induction, and \eq{\bm{v} = \nabla_{\bm{k}} \epsilon / \hbar} is the group velocity. We neglect Berry curvature effects and assume a single parabolic band with dispersion relation \eq{\epsilon = \hbar^2 k^2 / 2m}, where \eq{\hbar} is the reduced Planck constant and \eq{m} is the effective mass. Inserting these relations into Eq.~\eqref{eqb:boltzmann_equation} gives
\begin{equation} \label{eqb:boltzmann_equation_2}
    \pdv{f}{t} - \frac{e}{\hbar} ( \bm{E} + \bm{v} \cp \bm{B} ) \vdot \pdv{f}{\bm{k}} = -\frac{(f - f_0)}{\tau}.
\end{equation}
The electric current density, from which the relevant conductivities will be obtained, is
\begin{equation} \label{eqb:electric_current}
	\bm{J} = -e \!\! \int \!\! \frac{\dd^d k}{(2\pi)^d} \, \bm{v} f,
\end{equation}
To compute the relevant conductivities, we use the fields
\begin{align}
    \bm{E} &= \bm{E}_\mr{s} + \Re \left(\bm{\mc{E}}_\mr{w} e^{-i\omega t} \right),
    \\
    \bm{B} &= \bm{B}_\mr{s} + \Re \left(\bm{\mc{B}}_\mr{w} e^{-i\omega t} \right).
\end{align}
Motivated by the form of these, we use the \textit{ansatz}:
\begin{equation} \label{eqb:distribution_function_ansatz}
    f = f_0 + f_\mr{s} + \Re \left( f_\mr{w} e^{-i\omega t} \right) + \dots,
\end{equation}
where ``\eq{\dots}'' denotes higher-order harmonic corrections. Inserting Eq.~\eqref{eqb:distribution_function_ansatz} into Eq.~\eqref{eqb:electric_current} shows that the relevant conductivities come from terms in \eq{f_\mr{w}} that are proportional to either \eq{\bm{\mc{E}}_\mr{w}} or \eq{\bm{\mc{B}}_\mr{w}}. Specifically, we are only interested in terms that are linear in \eq{\mc{E}^i_\mr{w}}, \eq{\mc{E}^i_\mr{w} \mc{B}^j_\mr{s}}, or \eq{\mc{B}^i_\mr{w} \mc{E}^j_\mr{s}}. Inserting Eq.~\eqref{eqb:distribution_function_ansatz} into Eq.~\eqref{eqb:boltzmann_equation_2} gives
\begin{align} \label{eqb:boltzmann_eq_ins}
    0 &= \frac{f_\mr{s}}{\tau} - \frac{e}{\hbar} (\bm{E}_\mr{s} + \bm{v} \cp \bm{B}_\mr{s}) \vdot \pdv{(f_0 + f_\mr{s})}{\bm{k}} - \frac{e}{4 \hbar} (\bm{\mc{E}}_\mr{w} + \bm{v} \cp \bm{\mc{B}}_\mr{w}) \vdot \pdv{f^*_\mr{w}}{\bm{k}} - \frac{e}{4 \hbar} (\bm{\mc{E}}^*_\mr{w} + \bm{v} \cp \bm{\mc{B}}^*_\mr{w}) \vdot \pdv{f_\mr{w}}{\bm{k}}  \notag
    \\
    &+ \Re \left\{ e^{-i\omega t} \left[ (\tau^{-1} - i\omega) f_\mr{w} - \frac{e}{\hbar} (\bm{\mc{E}}_\mr{w} + \bm{v} \cp \bm{\mc{B}}_\mr{w}) \vdot \pdv{(f_0 + f_\mr{s})}{\bm{k}} - \frac{e}{\hbar} (\bm{E}_\mr{s} + \bm{v} \cp \bm{B}_\mr{s})  \vdot \pdv{f_\mr{w}}{\bm{k}} \right] \right\} + \dots.
\end{align}
The momentum-space gradient of the equilibrium distribution is \eq{\partial_{\bm{k}} f_0 = \hbar \bm{v} \partial_\epsilon f_0} and each order of \eq{\exp(i\omega t)} in Eq.~\eqref{eqb:boltzmann_eq_ins} must vanish separately, leading to
\begin{align}
    f_\mr{s} &= e\tau \frac{\partial f_0}{\partial \epsilon}  \bm{E}_\mr{s} \vdot \bm{v} + \frac{e\tau}{\hbar} \left[ (\bm{E}_\mr{s} + \bm{v} \cp \bm{B}_\mr{s}) \vdot \pdv{f_\mr{s}}{\bm{k}} + { \frac{1}{4}} (\bm{\mc{E}}_\mr{w} + \bm{v} \cp \bm{\mc{B}}_\mr{w}) \vdot \pdv{f^*_\mr{w}}{\bm{k}} + {\frac{1}{4}}(\bm{\mc{E}}^*_\mr{w} + \bm{v} \cp \bm{\mc{B}}^*_\mr{w}) \vdot \pdv{f_\mr{w}}{\bm{k}} \right] \dots, \label{eqb:f_s}
    \\
    f_\mr{w} &= \frac{e \tau}{\hbar(1 - i\omega \tau)} \left[ \hbar \pdv{f_0}{\epsilon} \bm{\mc{E}}_\mr{w} \vdot \bm{v} + (\bm{\mc{E}}_\mr{w} + \bm{v} \cp \bm{\mc{B}}_\mr{w}) \vdot \pdv{f_\mr{s}}{\bm{k}} +  (\bm{E}_\mr{s} + \bm{v} \cp \bm{B}_\mr{s}) \vdot \pdv{f_\mr{w}}{\bm{k}} \right] + \dots.
\end{align}
Inserting the recursive solution for \eq{f_\mr{s}} into \eq{f_\mr{w}} and then solving recursively for \eq{f_\mr{w}}, where using \begin{equation}
    \pdv{\bm{k}} \left[\pdv{f_0}{\epsilon} \bm{E}_\mr{s} \vdot \bm{v} \right] = \hbar \pdv[2]{f_0}{\epsilon} \bm{v} (\bm{E}_\mr{s} \vdot \bm{v}) + \frac{\hbar}{m} \pdv{f_0}{\epsilon} \bm{E}_\mr{s},
\end{equation}
we retain the relevant terms:
\begin{equation}
    f_\mr{w} = \frac{e \tau}{(1 - i\omega \tau)} \frac{\partial f_0}{\partial \epsilon} \bm{\mc{E}}_\mr{w} \vdot \bm{v} + \frac{e^2 \tau^2}{m(1 - i\omega \tau)} \frac{\partial f_0}{\partial \epsilon} (\bm{v} \cp \bm{\mc{B}}_\mr{w}) \vdot \bm{E}_\mr{s} + \frac{e^2 \tau^2}{m (1 - i\omega \tau)^2} \frac{\partial f_0}{\partial \epsilon} (\bm{v} \cp \bm{B}_\mr{s}) \vdot \bm{\mc{E}}_\mr{w} + \dots.
\end{equation}
The electric current follows from Eq.~\eqref{eqb:electric_current}, from which the conductivities are identified as
\begin{align}
	\sigma^{ij}_\mr{E}(\omega) &= \frac{e^2 \tau}{(1 - i\omega \tau)}  \!\! \int \!\! \frac{\dd^d k}{(2\pi)^d} \left( \!\!-\pdv{f_0}{\epsilon} \right) v^i v^j,  \label{eqb:sigma_e}
	\\
	\sigma^{ijk}_\mr{EB}(0, \omega) &= \frac{e^3 \tau^2}{m(1 - i\omega \tau)} \!\! \int \!\! \frac{\dd^d k}{(2\pi)^d} \left( \!\!-\pdv{f_0}{\epsilon} \right) v^i \varepsilon^{jlk} v^l, \label{eqb:sigma_eh}
	\\
	\sigma^{ijk}_\mr{EB}(\omega, 0) &= \frac{e^3 \tau^2}{m(1 - i\omega \tau)^2}  \!\! \int \!\! \frac{\dd^d k}{(2\pi)^d} \left( \!\!-\pdv{f_0}{\epsilon} \right) v^i \varepsilon^{jlk} v^l, \label{eqb:sigma_he}
\end{align}
where repeated superscripts denoting spatial components are implicitly summed over. Such integrals over momentum are described, for example, in Ref.~\cite{smith_1989}. Here, we perform them as
\begin{align} \label{eqb:momentum_integration}
    \int \!\! \frac{\dd^d k}{(2\pi)^d} \left( \!\!-\pdv{f_0}{\epsilon} \right) v^i v^j &= \! \int\limits^\infty_0 \!\! \dd \epsilon \left( \!\!-\pdv{f_0}{\epsilon} \right) D_d \langle v^i v^j  \rangle_\text{ang.} = \int\limits^\infty_0 \!\! \dd \epsilon \left( \!\!-\pdv{f_0}{\epsilon} \right) D_d \frac{\delta^{ij}2\epsilon}{dm} 
    \\
    &= \delta^{ij} \frac{2}{d m} \! \int\limits^\infty_0 \!\! \dd \epsilon \, f_0 \pdv{\epsilon} (D_d \epsilon) = \delta^{ij} \frac{1}{m}  \! \int\limits^\infty_0 \!\! \dd \epsilon \, f_0 D_d = \delta^{ij} \frac{n_d}{m}. \notag
\end{align}
In evaluating the integral, we first expressed the integral in terms of the \eq{d}-dimensional density of states \eq{D_d}, followed by performing the angular average \eq{\langle v^i v^j \rangle_\text{ang.} = \delta^{ij} v^2 / d}. Then, integrating by parts and applying the dimensional scaling of the density of states, \eq{D_d(\epsilon) \sim \epsilon^{d/2 - 1}}, yields the final expression in terms of the \eq{d}-dimensional density \eq{n_d}. The boundary terms \eq{f_0 D_d \epsilon|^\infty_0} from integrating by parts can be verified to vanish using the dimensional scaling of the density of states. Substituting Eq. \eqref{eqb:momentum_integration} into Eqs.~(\ref{eqb:sigma_e}, \ref{eqb:sigma_eh}) and using \eq{\bm{B} = \mu_0 \bm{H}} gives Eqs.~(7a, 7b) of the main text.

\section{Reflection matrix for 2D and semi-infinite systems}
In this section, we derive the reflection matrix for 2D and semi-infinite systems. The coordinate system and placement of the material are shown in Fig.~(1) of the main text. We employ a macroscopic formulation of Maxwell's equations suitable for optical frequencies presented in Ref. \cite{landau_1984}, which reads
\begin{align}   
    \div \bm{D} &= 0, \label{eqc:gauss_law}
    \\
    \div \bm{H} &= 0, \label{eqc:gauss_law_magnetism}
    \\
    \curl \bm{E} &= -\mu_0 \pdv{\bm{H}}{t}, \label{eqc:faradays_law}
    \\
    \curl \bm{H} &= \pdv{\bm{D}}{t}. \label{eqc:amperes_law}
\end{align}
The polarization \eq{\bm{P}} is in this formulation related to the total microscopic charge density via \eq{\rho = -\div \bm{P}}. The material response is described by either the polarization, electric current \eq{\bm{J}}, or electric displacement field \eq{\bm{D}}, since they are related through \eq{\bm{J} = \partial_t \bm{P}} and \eq{\bm{D} = \varepsilon_0 \bm{E} + \bm{P}}, where \eq{\bm{E}} is the electric field and \eq{\varepsilon_0} is the vacuum permittivity. Magnetic dipole responses are typically weak and negligible at optical frequencies \cite{landau_1984}, so the magnetic field \eq{\bm{H}} and the magnetic induction \eq{\bm{B}} are related by the vacuum permeability \eq{\mu_0} as \eq{\bm{H} = \bm{B} / \mu_0}. In the following sections, we determine the reflection matrix for the two-dimensional and semi-infinite material by solving the macroscopic Maxwell's equations in all regions and applying appropriate boundary conditions to obtain the reflected wave.

\subsection*{Above the material}
We begin with the region common to both the 2D and semi-infinite systems, corresponding to \eq{z < 0} in Fig.~(1) of the main text. The wave equation is obtained by taking the curl of Eq.~\eqref{eqc:faradays_law} and substituting Eq.~\eqref{eqc:amperes_law}, which gives
\begin{equation} \label{eqc:wave_equation}
	\curl (\curl \bm{E}_-) = - \mu_0 \pdv[2]{\bm{D}_-}{t}.
\end{equation}
The subscript ``\eq{-}'' denotes the \eq{z < 0} region, where we use the constitutive relation \eq{\bm{D}_- = \varepsilon_- \bm{E}_-} with a frequency-independent scalar dielectric constant \eq{\varepsilon_-}. Inserting an electromagnetic plane wave with wavevector \eq{\bm{k}_-} and angular frequency \eq{\omega} yields the dispersion relation \eq{k_- = \omega / v_- }, where the wave velocity is \eq{v_- = 1 / \sqrt{\mu_0 \varepsilon_-}}. We take the incident wave to propagate in the \eq{yz}-plane with wavevector \eq{\bm{k}_\mr{i} = k_- [\hat{\bm{y}} \sin(\phi_\mr{i}) + \hat{\bm{z}} \cos(\phi_\mr{i})]}, where \eq{\phi_\mr{i}} is the angle of incidence relative to the \eq{z}-axis, as shown in Fig.~(1) of the main text. At the \eq{z=0} interface, the in-plane components of the wavevector for all waves must match to satisfy the boundary conditions imposed by Maxwell's equations. Thus, the reflected wave has wavevector \eq{\bm{k}_\mr{r} = k_- [\hat{\bm{y}} \sin(\phi_\mr{i}) - \hat{\bm{z}} \cos(\phi_\mr{i})]}. We adopt the following \textit{ansatz} for the wave (\eq{\mr{w}}) part of the electric and magnetic fields in this region:
\begin{align}
	\bm{E}_{- \mr{w}}(\bm{r}, t) &= \Re \left[ \bm{\mc{E}}_\mr{i} e^{i(k^y_\mr{i} y + k^z_\mr{i} z - \omega t)} + \bm{\mc{E}}_\mr{r} e^{i(k^y_\mr{i} y - k^z_\mr{i} z - \omega t)} \right],
	\\
	\bm{H}_{- \mr{w}}(\bm{r}, t) &= \Re \left[ \bm{\mc{H}}_\mr{i} e^{i(k^y_\mr{i} y + k^z_\mr{i} z - \omega t)} + \bm{\mc{H}}_\mr{r} e^{i(k^y_\mr{i} y - k^z_\mr{i} z - \omega t)} \right].
\end{align}
Here, calligraphic symbols denote the complex amplitudes. The number of independent unknowns in this \textit{ansatz} can be reduced by applying Gauss's and Faraday's laws. Using Eq.~\eqref{eqc:gauss_law}, we obtain
\begin{align} 
	\mc{E}^z_\mr{i} &= - \tan(\phi_\mr{i}) \mc{E}^y_\mr{i}, \label{eqc:result_gauss_law1}
	\\
	\mc{E}^z_\mr{r} &= \tan(\phi_\mr{i}) \mc{E}^y_\mr{r}. \label{eqc:result_gauss_law2}
\end{align}
Applying Eq.~\eqref{eqc:faradays_law} together with Eqs.~(\ref{eqc:result_gauss_law1}, \ref{eqc:result_gauss_law2}) gives
\begin{align}
	\bm{\mc{H}}_\mr{i} &= \frac{1}{Z_-} \left[ -\hat{\bm{x}} \sec(\phi_\mr{i}) \mc{E}^y_\mr{i} + \hat{\bm{y}} \cos(\phi_\mr{i}) \mc{E}^x_\mr{i} - \hat{\bm{z}} \sin(\phi_\mr{i}) \mc{E}^x_\mr{i} \right],
	\\
	\bm{\mc{H}}_\mr{r} &= \frac{1}{Z_-} \left[ \hat{\bm{x}} \sec(\phi_\mr{i}) \mc{E}^y_\mr{r} - \hat{\bm{y}} \cos(\phi_\mr{i}) \mc{E}^x_\mr{r} - \hat{\bm{z}} \sin(\phi_\mr{i}) \mc{E}^x_\mr{r} \right],
\end{align}
where \eq{Z_- = \sqrt{\mu_0 / \varepsilon_-}} is the wave impedance.

\subsection*{Transformation to \eq{s}-\eq{p} basis}
We will later follow a similar procedure for the \eq{z > 0} region and use appropriate boundary conditions to determine the Kerr rotation. This will be carried out separately for the 2D and semi-infinite systems, yielding two matrices \eq{F} and \eq{G} that define the set of equations:
\begin{equation} \label{eqc:F_G_matrices}
    \begin{pmatrix}
        F_{xx} & F_{xy} 
        \\
        F_{yx} & F_{yy}
    \end{pmatrix}
    \begin{pmatrix}
        \mc{E}^x_\mr{r} 
        \\
        \mc{E}^y_\mr{r}
    \end{pmatrix}
    =
    \begin{pmatrix}
        G_{xx} & G_{xy}
        \\
        G_{yx} & G_{yy}
    \end{pmatrix}
    \begin{pmatrix}
        \mc{E}^x_\mr{i}
        \\
        \mc{E}^y_\mr{i}
    \end{pmatrix}.
\end{equation}
To express this in the perpendicular (\textit{senkrecht} or \eq{s}) and parallel (\eq{p}) polarization basis, we define the unit vectors \eq{\hat{\bm{s}}_\mr{i} = \hat{\bm{s}}_\mr{r} = \hat{\bm{x}}} and
\begin{align}
	\hat{\bm{p}}_\mr{i} &= \frac{\hat{\bm{k}}_\mr{i} \cp \hat{\bm{s}}_\mr{i}}{|\hat{\bm{k}}_\mr{i} \cp \hat{\bm{s}}_\mr{i}|} = \hat{\bm{y}} \cos(\phi_\mr{i}) - \hat{\bm{z}} \sin(\phi_\mr{i}), \label{eqc:parallell_incident}
	\\
	\hat{\bm{p}}_\mr{r} &= \frac{\hat{\bm{k}}_\mr{r} \cp \hat{\bm{s}}_\mr{r}}{| \hat{\bm{k}}_\mr{r} \cp \hat{\bm{s}}_\mr{r} |} = -\hat{\bm{y}} \cos(\phi_\mr{i}) - \hat{\bm{z}} \sin(\phi_\mr{i}). \label{eqc:parallell_reflected}
\end{align}
The \eq{s}-\eq{p} components of the electric field amplitudes follow as 
\begin{align}
    \mc{E}^s_\mr{i} &= \hat{\bm{s}}_\mr{i} \vdot \bm{\mc{E}}_\mr{i} = \mc{E}^x_\mr{i},
    \\
    \mc{E}^p_\mr{i} &= \hat{\bm{p}}_\mr{i} \vdot \bm{\mc{E}}_\mr{i} = \sec(\phi_\mr{i}) \mc{E}^y_\mr{i},
    \\
    \mc{E}^s_\mr{r} &= \hat{\bm{s}}_\mr{r} \vdot \bm{\mc{E}}_\mr{r} = \mc{E}^x_\mr{r},
    \\
    \mc{E}^p_\mr{r} &= \hat{\bm{p}}_\mr{r} \vdot \bm{\mc{E}}_\mr{r} = -\sec(\phi_\mr{i}) \mc{E}^y_\mr{r},
\end{align}
where Eqs.~(\ref{eqc:result_gauss_law1}, \ref{eqc:result_gauss_law2}) have been used to eliminate \eq{\mc{E}^z_\mr{i}} and \eq{\mc{E}^z_\mr{r}}. In matrix form, we define the basis rotation matrices \eq{R_\mr{i} = \text{diag}[1,\sec(\phi_\mr{i})]} and \eq{R_\mr{r} = \text{diag}[1,-\sec(\phi_\mr{i})]}, such that rotating Eq.~\eqref{eqc:F_G_matrices} to the \eq{s}–\eq{p} basis as
\begin{equation}
	F (R^{-1}_\mr{r} R_\mr{r}) \begin{pmatrix} \mc{E}^x_\mr{r} \\ \mc{E}^y_\mr{r} \end{pmatrix} 
	= 
	G (R^{-1}_\mr{i} R_\mr{i}) \begin{pmatrix} \mc{E}^x_\mr{i} \\ \mc{E}^y_\mr{i} \end{pmatrix}
	\iff
	\begin{pmatrix} \mc{E}^s_\mr{r} \\ \mc{E}^p_\mr{r} \end{pmatrix}
	=
	r \begin{pmatrix} \mc{E}^s_\mr{i} \\ \mc{E}^p_\mr{i} \end{pmatrix},
\end{equation}
gives the reflection matrix
\begin{equation} \label{eqc:reflection_matrix}
	r = R_\mr{r} F^{-1} G R^{-1}_\mr{i}. 
\end{equation}
The matrix elements are
\begin{align}
    r_{ss} &= \frac{1}{\det(F)} (F_{yy} G_{xx} - F_{xy} G_{yx}), 
    &
    r_{sp} &= \frac{\cos(\phi_\mr{i})}{\det(F)} (F_{yy}G_{xy} - F_{xy} G_{yy}),
    \\
    r_{ps} &= \frac{\sec(\phi_\mr{i})}{\det(F)} (F_{yx} G_{xx} - F_{xx} G_{yx} ), 
    &
    r_{pp} &= \frac{1}{\det(F)} (F_{yx} G_{xy} - F_{xx} G_{yy}).
\end{align}

\subsection*{2D material}
We now consider a 2D material confined to the \eq{z = 0} plane. In the \eq{z > 0} region, we use the constitutive relation \eq{\bm{D}_+ = \varepsilon_+ \bm{E}_+}, where the subscript ``\eq{+}'' denotes the region and \eq{\varepsilon_+} is a frequency-independent scalar dielectric constant. The wave equation in Eq.~\eqref{eqc:wave_equation} also applies in this region by interchanging the subscripts ``\eq{-}''~\eq{\to}~``\eq{+}'', such that the dispersion relation is given by \eq{k_+ = \omega / v_+} with wave velocity \eq{v_+ = 1 / \sqrt{\varepsilon_+ \mu_0}}. We write the wavevector of the transmitted \eq{(\mr{t})} wave as \eq{\bm{k}_\mr{t} = k_+ [\hat{\bm{y}} \sin(\phi_\mr{t}) + \hat{\bm{z}} \cos(\phi_\mr{t})]}. The transmitted angle \eq{\mr{\phi}_\mr{t}} is fixed by continuity of the in-plane components of the wavevectors for the incident, reflected, and transmitted waves at the \eq{z = 0} interface:
\begin{equation} \label{eqc:snells_law}
	k_- \sin(\phi_\mr{i}) = k_+ \sin(\phi_\mr{t}) \iff \frac{\sin(\phi_\mr{i})}{\sin(\phi_\mr{t})} = \frac{v_-}{v_+} = \sqrt{\frac{\varepsilon_+}{\varepsilon_-}} = \frac{Z_-}{Z_+},
\end{equation}
where the wave impedance in the \eq{z > 0} region is \eq{Z_+ = \sqrt{\mu_0/\varepsilon_+}}. We use the following \textit{ansatz} for the wave part of the electric and magnetic fields for the \eq{z > 0} region,
\begin{align}
	\bm{E}_{+ \mr{w}}(\bm{r}, t) &= \Re \left[ \bm{\mc{E}}_\mr{t} e^{i(k^y_\mr{t} y + k^z_\mr{t} z - \omega t)} \right],
	\\
	\bm{H}_{+ \mr{w}}(\bm{r}, t) &= \Re \left[ \bm{\mc{H}}_\mr{t} e^{i(k^y_\mr{t} y + k^z_\mr{t} z - \omega t)} \right].
\end{align}
As before, the number of independent unknowns is reduced using Gauss's and Faraday's laws. Equation~\eqref{eqc:gauss_law} gives
\begin{equation} \label{eqc:gauss_law_result2}
	\mc{E}^z_\mr{t} = -\tan(\phi_\mr{t}) \mc{E}^y_\mr{t},
\end{equation}
and Eq.~\eqref{eqc:faradays_law} combined with Eq.~\eqref{eqc:gauss_law_result2} yields
\begin{equation}
	\bm{\mc{H}}_\mr{t} = \frac{1}{Z_+} \left[ -\hat{\bm{x}} \sec(\phi_\mr{t}) \mc{E}^y_\mr{t} + \hat{\bm{y}} \cos(\phi_\mr{t}) \mc{E}^x_\mr{t} - \hat{\bm{z}} \sin(\phi_\mr{t}) \mc{E}^x_\mr{t} \right].
\end{equation}
The boundary condition at \eq{z=0} for the electric field follows from Eq.~\eqref{eqc:faradays_law}, where we assume no delta function contribution to \eq{\bm{H}}. This gives
\begin{align} 
	\mc{E}^x_\mr{t} &= \mc{E}^x_\mr{i} + \mc{E}^x_\mr{r}, \label{eqc:boundary_condition_E1}
	\\
	\mc{E}^y_\mr{t} &= \mc{E}^y_\mr{i} + \mc{E}^y_\mr{r}. \label{eqc:boundary_condition_E2}
\end{align}
The boundary condition for the magnetic field is presented in Eq.~(8) of the main text and reads, in component form,
\begin{align} 
	\mc{H}^y_\mr{i} + \mc{H}^y_\mr{r} - \mc{H}^y_\mr{t} &= \mc{K}_\mr{w}^x, \label{eqc:boundary_condition_H1}
	\\
	\mc{H}^x_\mr{t} - \mc{H}^x_\mr{i} - \mc{H}^x_\mr{r} &= \mc{K}_\mr{w}^y. \label{eqc:boundary_condition_H2}
\end{align}
The wave part of the surface current \eq{\bm{K}_\mr{w}} is related to the averaged electric and magnetic fields as detailed in Eq.~(8) of the main text:
\begin{align}
    \bm{K}_\mr{w} = \Re \left[ \sigma_\mr{e}(\omega) \frac{1}{2} (\bm{E}_{-\mr{w}}|_{z = 0^-} +  \bm{E}_{+\mr{w}}|_{z = 0^+}) + \sigma_\mr{eh}(0, \omega) \bm{E}_0 \cp \frac{1}{2}(\bm{H}_{-\mr{w}}|_{z = 0^-} +  \bm{H}_{+\mr{w}}|_{z = 0^+} ) \right],
\end{align}
where the conductivities are given in Eqs.~(\ref{eqb:sigma_e}, \ref{eqb:sigma_eh}) and include the 2D number density due to the relation between the surface current \eq{\bm{K}} and current, \eq{\bm{J} = \delta(z) \bm{K}}. From here on, we omit the explicit \eq{\omega} dependence of the conductivities for notational convenience. Inserting the \textit{ansatz} for \eq{\bm{E}_{\pm \mr{w}}} and \eq{\bm{H}_{\pm \mr{w}}} into Eqs.~(\ref{eqc:boundary_condition_H1}, \ref{eqc:boundary_condition_H2}) combined with using Eqs.~(\ref{eqc:boundary_condition_E1}, \ref{eqc:boundary_condition_E2}, \ref{eqc:snells_law}) gives
\begin{align}
	\frac{\cos(\phi_\mr{i})}{Z_-} (\mc{E}^x_\mr{i} - \mc{E}^x_\mr{r}) - \frac{\cos(\phi_\mr{t})}{Z_+} (\mc{E}^x_\mr{i} + \mc{E}^x_\mr{r}) &= \sigma_\mr{e} (\mc{E}^x_\mr{i} + \mc{E}^x_\mr{r}) - \sigma_\mr{eh} E^y_0 \frac{\sin(\phi_\mr{i})}{Z_-} (\mc{E}^x_\mr{i} + \mc{E}^x_\mr{r}),
	\\
	\frac{\sec(\phi_\mr{i})}{Z_-} (\mc{E}^y_\mr{i} - \mc{E}^y_\mr{r}) - \frac{\sec(\phi_\mr{t})}{Z_+} (\mc{E}^y_\mr{i} + \mc{E}^y_\mr{r}) &= \sigma_\mr{e} (\mc{E}^y_\mr{i} + \mc{E}^y_\mr{r}) + \sigma_\mr{eh} E^x_0 \frac{\sin(\phi_\mr{i})}{Z_-} (\mc{E}^x_\mr{i} + \mc{E}^x_\mr{r}).
\end{align}
These two equations can be written in matrix form, with \eq{F} and \eq{G} defined in Eq.~\eqref{eqc:F_G_matrices}, where the matrix elements for the 2D material are
\begin{align}
    F_{xx} &= 1 + \frac{Z_- \cos(\phi_\mr{t})}{Z_+ \cos(\phi_\mr{i})} + Z_- \sigma_\mr{e} \sec(\phi_\mr{i}) - \sigma_\mr{eh} E^y_0 \tan(\phi_\mr{i}), & G_{xx} &= 2 - F_{xx},
    \\
    F_{xy} &= 0, & G_{xy} &= 0,
    \\
    F_{yx} &= \sigma_\mr{eh} E^x_0 \sin(\phi_\mr{i}) \cos(\phi_\mr{i}), & G_{yx} &= -F_{yx}, 
    \\
    F_{yy} &= 1 + \frac{Z_- \cos(\phi_\mr{i})}{Z_+ \cos(\phi_\mr{t})} + Z_- \sigma_\mr{e} \cos(\phi_\mr{i}), & G_{yy} &= 2 - F_{yy}.
\end{align}
Using Eq.~\eqref{eqc:reflection_matrix} yields the reflection matrix for the 2D material in the \eq{s}-\eq{p} basis with matrix elements:
\begin{align} 
    r_{ss} &= \frac{(2-F_{xx})}{F_{xx}}, 
    &
    r_{sp} &= 0, \label{eqc:2d_reflection1}
    \\
    r_{ps} &= \frac{2\sec(\phi_\mr{i})F_{yx}}{F_{xx} F_{yy}}, 
    &
    r_{pp} &= \frac{(F_{yy}-2)}{F_{yy}}. \label{eqc:2d_reflection2}
\end{align}

\subsection*{Semi-infinite material}
We now consider a semi-infinite material confined to the \eq{z \geq 0} region. In this region, we have \eq{\bm{D}_{+\mr{w}} = \varepsilon_0 \bm{E}_{+\mr {w}} + \bm{P}_{+\mr{w}}}, which combined with \eq{\bm{J} = \partial_t \bm{P}} and the constitutive relation specified in Eq.~(6) of the main text, gives
\begin{equation}
	\bm{D}_{+ \mr{w}} = \Re \left[\varepsilon_\mr{e}(\omega) \bm{E}_{+ \mr{w}} + \varepsilon_\mr{eh}(0, \omega) \bm{E}_0 \cp \bm{H}_{+ \mr{w}} \right].
\end{equation}
The subscript ``\eq{+}'' now denotes the \eq{z \geq 0} region, and the permittivities are related to the conductivities as
\begin{align}
	\varepsilon_\mr{e}(\omega) &= \varepsilon_0 + \frac{i}{\omega} \sigma_\mr{e}(\omega),
	\\
	\varepsilon_\mr{eh}(0, \omega) &= \frac{i}{\omega} \sigma_\mr{eh}(0, \omega).
\end{align}
Hereafter, we omit the explicit \eq{\omega} dependence of the permittivities for notational convenience. Since the in-plane components of the wavevectors for the incident, reflected, and transmitted waves must match at \eq{z=0}, we use \eq{\bm{k}_\mr{t} = \hat{\bm{y}} k_- \sin(\phi_\mr{i}) + \hat{\bm{z}} k^z_+}, where transmitted (\eq{\mr{t}}) now refers to the wave inside the material. Thus, we need to find the dispersion relation for the normal component of the wavevector inside the material. We use a plane wave \textit{ansatz} for the wave part of the electric and magnetic fields,
\begin{align} 
	\bm{E}_{+ \mr{w}}(\bm{r}, t) &= \Re \left[ \bm{\mc{E}}_\mr{t} e^{i(k^y_\mr{t} y + k^z_\mr{t} z - \omega t)} \right], \label{eqc:ansatz_inside_E}
	\\
	\bm{H}_{+ \mr{w}}(\bm{r}, t) &= \Re \left[ \bm{\mc{H}}_\mr{t} e^{i(k^y_\mr{t} y + k^z_\mr{t} z - \omega t)} \right]. \label{eqc:ansatz_inside_B}
\end{align}
The wave equation inside the material is given in Eq.~\eqref{eqc:wave_equation} with substituting the subscripts ``\eq{-}''~\eq{\to}~``\eq{+}''. Inserting Eqs.~(\ref{eqc:ansatz_inside_E}, \ref{eqc:ansatz_inside_B}) into the wave equation gives
\begin{equation}
	\bm{k}_\mr{t} \cp (\bm{k}_\mr{t} \cp \bm{\mc{E}}_\mr{t}) + \mu_0 \omega^2 \left( \varepsilon_\mr{e} \bm{\mc{E}}_\mr{t} + \varepsilon_\mr{eh} \bm{E}_0 \cp \bm{\mc{H}}_\mr{t} \right) = 0.
\end{equation}
We write the magnetic amplitude in terms of the electric amplitude by using Eq.~\eqref{eqc:faradays_law}, which gives \eq{\mu_0 i\omega \bm{\mc{H}}_\mr{t} = i \bm{k}_\mr{t} \cp \bm{\mc{E}}_\mr{t}}. Substituting this into the above equation gives the matrix equation \eq{S \bm{\mc{E}}_\mr{t} = 0}, where
\begin{equation}
	S^{ij} = k^i_\mr{t} k^j_\mr{t} + \omega \varepsilon_\mr{eh} k^i_\mr{t} E^j_0 - \delta^{ij} \left( k^2_\mr{t} - \mu_0 \omega^2 \varepsilon_\mr{e} + \omega \varepsilon_\mr{eh} \bm{E}_0 \vdot \bm{k}_\mr{t} \right). \label{eqc:matrix_s}
\end{equation}
The allowed dispersion relations are found by requiring \eq{\det(S) = 0}, which gives
\begin{equation}
    \mu_0 \omega^2 \varepsilon_\mr{e} \left[-k^2_- \sin^2(\phi_\mr{i}) - (k^z_+)^2 +  \mu_0 \omega^2 \varepsilon_\mr{e} - \omega \varepsilon_\mr{eh} E^y_0 k_- \sin(\phi_\mr{i}) \right]^2 = 0.
\end{equation}
We solve this equation for \eq{k^z_+} and find the dispersion relation:
\begin{equation}
    k^z_+ = \frac{\omega}{v_+} \sqrt{1 - \frac{Z^2_+}{Z^2_-}\sin^2(\phi_\mr{i}) - \frac{\varepsilon_\mr{eh}E^y_0 }{Z_- \varepsilon_\mr{e}}\sin(\phi_\mr{i})},
\end{equation}
where we defined the frequency-dependent wave velocity \eq{v_+ = 1 / \sqrt{\mu_0 \varepsilon_\mr{e}}} and impedance \eq{Z_+ = \sqrt{\mu_0 / \varepsilon_\mr{e}}} inside the material. Note that only the \eq{y}-component of the static electric field enters the dispersion relation, and it appears with the ratio \eq{\varepsilon_\mr{eh} / \varepsilon_\mr{e}}, which is assumed to be much smaller than the preceding term proportional to \eq{1 / \varepsilon_\mr{e}}. In analogy to the 2D material, we define the two variables:
\begin{align}
    \sin_\mr{t} &= \frac{Z_+}{Z_-}\sin(\phi_\mr{i}), \label{eqc:sint}
    \\
    \cos_\mr{t} &= \sqrt{1 - \sin^2_\mr{t} - \frac{\varepsilon_\mr{eh}E^y_0}{Z_- \varepsilon_\mr{e}}\sin(\phi_\mr{i})}, \label{eqc:cost}
\end{align}
such that the transmitted wavector can be written as \eq{\bm{k}_\mr{t} = k_+ [\hat{\bm{y}} \sin_\mr{t} + \hat{\bm{z}} \cos_\mr{t}]} with \eq{k_+ = \omega / v_+}. We emphasize that \eq{\sin_\mr{t}} and \eq{\cos_\mr{t}} are here treated as independent variables that cannot be inverted and lack the geometric properties of standard trigonometric functions, because the optical permittivities \eq{\varepsilon_\mr{e}} and \eq{\varepsilon_\mr{eh}}, and hence impedance \eq{Z_+} and wave velocity \eq{v_+}, are in general complex-valued. The trigonometric identity \eq{\sin_\mr{t}^2 + \cos^2_\mr{t} = 1} is recovered only in the special case of \eq{E^y_0 = 0}. However, as hinted above and confirmed later, \eq{E^y_0} contributes only minor corrections to the Kerr rotation, and we later show that setting \eq{E^y_0 = 0} substantially simplifies the expressions using the trigonometric identity. As before, the number of independent unknowns in the \textit{ansatz} can be reduced by applying Faraday's and Gauss's laws. Using Eq.~\eqref{eqc:faradays_law}, we obtain
\begin{equation} \label{eqc:half_infinite_magnetic}
	\bm{\mc{H}}_\mr{t} = \frac{1}{Z_+} \left[ \hat{\bm{x}}(\sin_\mr{t} \mc{E}^z_\mr{t} - \cos_\mr{t} \mc{E}^y_\mr{t}) + \hat{\bm{y}} \cos_\mr{t} \mc{E}^x_\mr{t} - \hat{\bm{z}} \sin_\mr{t} \mc{E}^x_\mr{t} \right],
\end{equation}
and Eq.~\eqref{eqc:gauss_law} combined with Eq.~\eqref{eqc:half_infinite_magnetic} yields
\begin{equation} \label{eqc:half_infinite_electric_z}
\begin{split}
	&0 = \bm{k}_\mr{t} \vdot \bm{\mc{D}}_{+ \mr{w}} = \sin_\mr{t} \left( \varepsilon_\mr{e} \mc{E}^y_\mr{t} - \varepsilon_\mr{eh} E^x_0 \mc{H}^z_\mr{t} \right) + \cos_\mr{t} \left[ \varepsilon_\mr{e}\mc{E}^z_\mr{t} + \varepsilon_\mr{eh} \left( E^x_0 \mc{H}^y_\mr{t} - E^y_0 \mc{H}^x_\mr{t} \right) \right]
    \\
    &\iff  \mc{E}^z_\mr{t} = \frac{1}{\left(1 - \frac{\sin_\mr{t} \varepsilon_\mr{eh} E^y_0}{Z_+ \varepsilon_\mr{e}}\right)} 
    \left[ - \mc{E}^x_\mr{t} \frac{\varepsilon_\mr{eh} E^x_0}{\cos_\mr{t} \varepsilon_\mr{e} Z_+} \left(\sin^2_\mr{t} + \cos^2_\mr{t} \right) - \mc{E}^y_\mr{t} \tan_\mr{t} \left( 1 + \frac{\cos^2_\mr{t} \varepsilon_\mr{eh} E^y_0}{\sin_\mr{t} Z_+ \varepsilon_\mr{e} } \right)
    \right],
\end{split}
\end{equation}
where \eq{\tan_\mr{t} = \sin_\mr{t} / \cos_\mr{t}}. The boundary conditions at \eq{z = 0} for the electric and magnetic fields, derived from Eqs.~(\ref{eqc:faradays_law},~\ref{eqc:amperes_law}), are
\begin{align}
    \mc{E}^x_\mr{t} &= \mc{E}^x_\mr{i} + \mc{E}^x_\mr{r}, \label{eqc:half_infinite_electric_continuity1}
    \\
    \mc{E}^y_\mr{t} &= \mc{E}^y_\mr{i} + \mc{E}^y_\mr{r}. \label{eqc:half_infinite_electric_continuity2}
    \\
    \mc{H}^x_\mr{t} &= \mc{H}^x_\mr{i} + \mc{H}^x_\mr{r}, \label{eqc:half_infinite_magnetic_continuity1}
    \\
    \mc{H}^y_\mr{t} &= \mc{H}^y_\mr{i} + \mc{H}^y_\mr{r}, \label{eqc:half_infinite_magnetic_continuity2}
\end{align}
where we assumed no special surface effects appropriate for a bulk material. 
Substituting Eqs.~(\ref{eqc:half_infinite_magnetic}, \ref{eqc:half_infinite_electric_z}, \ref{eqc:half_infinite_electric_continuity1}, \ref{eqc:half_infinite_electric_continuity2}) into Eq.~\eqref{eqc:half_infinite_magnetic_continuity2} results in
\begin{equation}
    \frac{\cos_\mr{t}}{Z_+} (\mc{E}^x_\mr{i} + \mc{E}^x_\mr{r}) = \frac{\cos(\phi_\mr{i})}{Z_-} (\mc{E}^x_\mr{i} - \mc{E}^x_\mr{r}).
\end{equation}
A similar substitution into Eq.~\eqref{eqc:half_infinite_magnetic_continuity1} gives
\begin{equation}
    - \frac{\tan_\mr{t}}{Z_+} (\mc{E}^x_\mr{i} + \mc{E}^x_\mr{r}) \frac{\varepsilon_\mr{eh} E^x_0(\sin^2_\mr{t} + \cos^2_\mr{t})}{(Z_+ \varepsilon_\mr{e} - \sin_\mr{t} \varepsilon_\mr{eh} E^y_0)} - \frac{1}{Z_+ \cos_\mr{t}} (\mc{E}^y_\mr{i} + \mc{E}^y_\mr{r})\left[ \cos^2_\mr{t} + \sin^2_\mr{t} \left(\frac{Z_+ \varepsilon_\mr{e} \sin_\mr{t} + \cos^2_\mr{t} \varepsilon_\mr{eh} E^y_0}{Z_+ \varepsilon_\mr{e} \sin_\mr{t} - \sin^2_\mr{t} \varepsilon_\mr{eh} E^y_0} \right) \right] = \frac{\sec(\phi_\mr{i})}{Z_-} (\mc{E}^y_\mr{r} - \mc{E}^y_\mr{i}),
\end{equation}
These two equations can be written in matrix form, with \eq{F} and \eq{G} defined in Eq.~\eqref{eqc:F_G_matrices}, where the matrix elements are
\begin{align}
    F_{xx} &= 1 + \frac{Z_- \cos_\mr{t}}{Z_+ \cos(\phi_\mr{i})}, & G_{xx} &= 2 - F_{xx},
    \\
    F_{xy} &= 0, & G_{xy} &= F_{xy},
    \\
    F_{yx} &= \frac{Z_- \cos(\phi_\mr{i}) \tan_\mr{t} \varepsilon_\mr{eh} E^x_0 (\sin^2_\mr{t} + \cos^2_\mr{t})}{Z_+ (Z_+ \varepsilon_\mr{e} - \sin_\mr{t} \varepsilon_\mr{eh} E^y_0)}, & G_{yx} &= - F_{yx},
    \\
    F_{yy} &= 1 + \frac{Z_- \cos(\phi_\mr{i})}{Z_+ \cos_\mr{t}} \left[\cos^2_\mr{t} + \sin^2_\mr{t} \left(\frac{Z_+ \varepsilon_\mr{e} \sin_\mr{t} + \cos^2_\mr{t} \varepsilon_\mr{eh} E^y_0}{Z_+ \varepsilon_\mr{e} \sin_\mr{t} - \sin^2_\mr{t} \varepsilon_\mr{eh} E^y_0} \right) \right], & G_{yy} &= 2 - F_{yy}.
\end{align}
Using Eq.~\eqref{eqc:reflection_matrix} yields the reflection matrix for the semi-infinite material in the \eq{s}-\eq{p} basis, with matrix elements
\begin{align} 
    r_{ss} &= \frac{(2-F_{xx})}{F_{xx}}, 
    &
    r_{sp} &= 0, \label{eqc:semi_infinite_reflection1}
    \\
    r_{ps} &= \frac{2\sec(\phi_\mr{i})F_{yx}}{F_{xx} F_{yy}}, 
    &
    r_{pp} &= \frac{(F_{yy}-2)}{F_{yy}}. \label{eqc:semi_infinite_reflection2}
\end{align}

\section{Kerr rotation and material estimates}
In this section, we provide the full analytical expressions for the Kerr rotation for a 2D material and a semi-infinite bulk. We also list the material parameters used in the estimates reported in the main text.

\subsection*{Material parameters}
The wavelength of the incident wave is set to \eq{\lambda = 800 \; \si{\nano\meter}}. The material parameters used to evaluate the Drude conductivity \eq{\sigma_\mr{e}} and the unconventional Hall conductivity \eq{\sigma_\mr{eh}} are given in Tab.~\ref{tabd:table}.
\begin{center}
\renewcommand{\arraystretch}{1.5}
\begin{table}[h]
\begin{tabular}{|c|c|c|c|}
\hline
Material & Carrier density & Relaxation time \eq{[\si{\second}]} & Effective mass \eq{[m_e]} \\
\hline
2D gallium arsenide (\eq{\mr{GaAs}}) & \eq{2 \times 10^{11} \si{\per\square\centi\metre}} & \eq{1 \times 10^{-12}} & \eq{0.067}\\
\hline
2D molybdenum disulfide (\eq{\mr{MoS_2}}) & \eq{1 \times 10^{12} \si{\per\square\centi\metre}}  & \eq{6.4 \times 10^{-13}} & \eq{0.54}\\
\hline
Copper (\eq{\mr{Cu}}) & \eq{8.5 \times 10^{22} \si{\per\cubic\centi\metre}} & \eq{1.9 \times 10^{-14}} & \eq{1} \\
\hline
Aluminum (\eq{\mr{Al}}) & \eq{1.8 \times 10^{23} \si{\per\cubic\centi\metre}} & \eq{6 \times 10^{-15}} & \eq{1} \\
\hline
Titanium (\eq{\mr{Ti}}) & \eq{1.1 \times 10^{23} \si{\per\cubic\centi\metre}} & \eq{1.8 \times 10^{-15}} & \eq{1} \\
\hline
\end{tabular}
\caption{Material parameters. For \eq{\mr{GaAs}}, we use the effective mass and density reported in Ref.~\cite{chung_2021} and estimate the relaxation time assuming a modest-mobility sample. For \eq{\mr{MoS_2}}, we use the effective mass reported in Ref.~\cite{kadantsev_2012} and the carrier density and mobility reported in Ref.~\cite{wang_2012}. For \eq{\mr{Cu}}, \eq{\mr{Ti}}, and \eq{\mr{Al}}, we estimate the carrier densities based on the mass densities and the molar mass \cite{handbook_2021}, where we use one conduction electron for \eq{\mr{Cu}}, two for \eq{\mr{Ti}}, and three for \eq{\mr{Al}}. The relaxation times are computed using the Drude conductivities reported in Ref. \cite{handbook_2021} at room temperature.}
\end{table}
\renewcommand{\arraystretch}{1.0}
\label{tabd:table}
\end{center}

\subsection*{2D material}
Using the reflection matrix for the 2D material with matrix elements given in Eqs.~(\ref{eqc:2d_reflection1}, \ref{eqc:2d_reflection2}), we compute the Kerr angles: \eq{\Theta_p = r_{sp} / r_{pp} = 0} and
\begin{align}
    \Theta_s &= \frac{r_{ps}}{r_{ss}} = \frac{2\sigma_\mr{eh} E^x_0 \sin(\phi_\mr{i})}{\left[ 1 + \frac{Z_- \cos(\phi_\mr{i})}{Z_+ \cos(\phi_\mr{t})} + \sigma_\mr{e} Z_- \cos(\phi_\mr{i}) \right] \left[ 1 - \frac{Z_- \cos(\phi_\mr{t}) }{Z_+
    \cos(\phi_\mr{i})} - \sigma_\mr{e} Z_- \sec(\phi_\mr{i}) + \sigma_\mr{eh} E^y_0 \tan(\phi_\mr{i}) \right]}. \label{eqd:complex_kerr_angle_2d}
\end{align}
Here, \eq{E^y_0} enters in combination with the unconventional response \eq{\sigma_\mr{eh}} in the denominator. We thus find that \eq{E^y_0} has only a minor effect on the Kerr response, since \eq{\sigma_\mr{eh}} was assumed to be much smaller than the linear response \eq{\sigma_\mr{e}}. Equation~(9) of the main text is obtained by setting \eq{E^y_0 = 0} and defining the refractive index \eq{n_\pm = \sqrt{\varepsilon_\pm}}.

\begin{figure}[h]
    \centering
    \includegraphics[width=\linewidth]{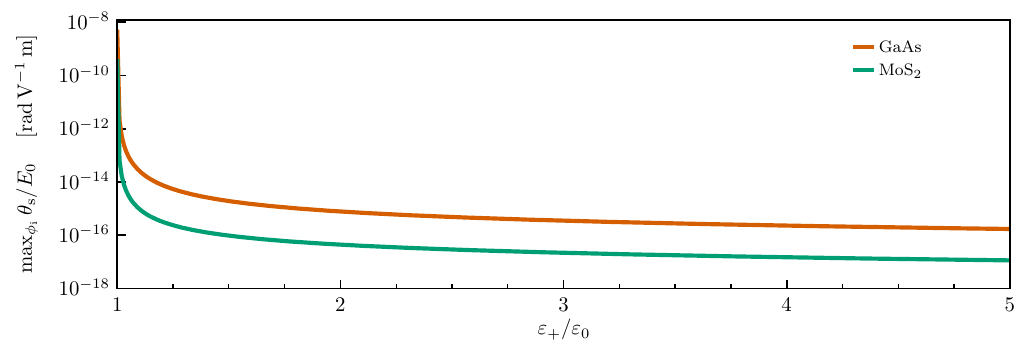}
    \caption{Maximum Kerr rotation angle per unit static electric field for the 2D materials \eq{\mr{GaAs}} and \eq{\mr{MoS_2}}, obtained by scanning over angles of incidence \eq{\phi_\mr{i} \in (0, \pi / 2)}, vs. dielectric constant below the material \eq{(z > 0)}. The static electric field is \eq{\bm{E}_0 = \hat{\bm{x}} E_0} and the wavelength of the incident wave is \eq{\lambda = 800 \; \si{\nano\meter}}}
    \label{figd:2D_media}
\end{figure}

We also investigate how asymmetric surrounding media (\eq{\varepsilon_- \neq \varepsilon_+}), modeling a substrate, influence the rotation magnitudes using Eq. \eqref{eqd:complex_kerr_angle_2d}. This is found to strongly suppress the Kerr rotation angle, as presented in Fig.~\ref{figd:2D_media}.

For \eq{\bm{E}_0 = \hat{\bm{x}} E_0} and a homogeneous medium surrounding the material, specified by \eq{\varepsilon_\pm = \varepsilon}, \eq{Z_\pm = \sqrt{\mu_0 / \varepsilon}}, and \eq{\phi_\mr{i} = \phi_\mr{t}}, Eq.~\eqref{eqd:complex_kerr_angle_2d} reduces to  
\begin{equation} \label{eqd:2d_equal_media}
	\Theta_s = \frac{-\sigma_\mr{eh} E_0 \sin(2 \phi_\mr{i})}{\sigma_\mr{e} Z \left[2 + \sigma_\mr{e} Z \cos(\phi_\mr{i}) \right]}. 
\end{equation}
The ratio \eq{\sigma_\mr{eh} / \sigma_\mr{e}}, where we remember that \eq{\sigma_\mr{e} = \sigma_\mr{e}(\omega)} and \eq{\sigma_\mr{eh} = \sigma_\mr{eh}(0, \omega)}, is from Eqs.~(\ref{eqb:sigma_e}, \ref{eqb:sigma_eh}) seen to be purely real. Therefore, we decompose the linear conductivity into its real and imaginary part as \eq{\sigma_\mr{e} = \sigma'_\mr{e} + i \sigma''_\mr{e}} with \eq{\sigma'_\mr{e}, \sigma''_\mr{e} \in \mathbb{R}}, and expand the term enclosed by brackets in the denominator as
\begin{equation}
	\frac{1}{2 + \sigma_\mr{e} Z \cos(\phi_\mr{i})} = \frac{2 + \sigma'_\mr{e} Z \cos(\phi_\mr{i}) - i \sigma''_\mr{e} Z \cos(\phi_\mr{i})}{\left[ 2 + \sigma'_\mr{e} Z \cos(\phi_\mr{i}) \right]^2 + \left[ \sigma''_\mr{e} Z \cos(\phi_\mr{i}) \right]^2}.
\end{equation}
We identify the Kerr rotation angle as the real part of Eq.~\eqref{eqd:2d_equal_media}, where in the limit of \eq{Z\sigma'_\mr{e} \ll 2} and \eq{Z\sigma''_\mr{e} \ll 2} we find
\begin{equation}
	\theta_s = - \frac{\sigma_\mr{eh}}{2 \sigma_\mr{e} Z} E_0 \sin(2 \phi_\mr{i}).
\end{equation}
Inserting the conductivities computed from the Boltzmann transport equation in Eqs.~(\ref{eqb:sigma_e}, \ref{eqb:sigma_eh}) gives Eq.~(1) of the main text. The transmission coefficient for linear 2D materials is \eq{T = |1 + \sigma_\mr{e}Z/2|^{-2}} \cite{li_2018}, such that the limits \eq{\sigma'_\mr{e} Z \ll 2} and \eq{\sigma''_\mr{e} Z \ll 2} correspond to the transparent limit. For the 2D materials considered with \eq{\varepsilon = \varepsilon_0}, we compute \eq{\sigma_\mr{e} Z \sim 10^{-7} + i 10^{-4}} for \eq{\mr{GaAs}} and \eq{\sigma_\mr{e} Z \approx 10^{-7} + i 10^{-4}} for \eq{\mr{MoS_2}}, which confirms the transparent regime for the 2D materials considered.

\subsection*{Semi-infinite material}
Using the reflection matrix for the semi-infinite material with matrix elements given in Eqs.~(\ref{eqc:semi_infinite_reflection1}, \ref{eqc:semi_infinite_reflection2}), we compute the Kerr angles: \eq{\Theta_p = r_{sp} / r_{pp} = 0} and
\begin{align}
    \Theta_s = \frac{r_{ps}}{r_{ss}} &= \frac{2Z_- \tan_t \varepsilon_\mr{eh} E^x_0 (\sin^2_\mr{t} + \cos^2_\mr{t})}{Z_+(Z_+ \varepsilon_\mr{e} - \sin_\mr{t} \varepsilon_\mr{eh} E^y_0)} \left[1 - \frac{Z_- \cos_\mr{t}}{Z_+ \cos(\phi_\mr{i})} \right]^{-1} \left\{1 + \frac{Z_- \cos(\phi_\mr{i})}{Z_+ \cos_\mr{t}} \left[\cos^2_\mr{t} + \sin^2_\mr{t} \left(\frac{Z_+ \varepsilon_\mr{e} \sin_\mr{t} + \cos^2_\mr{t} \varepsilon_\mr{eh} E^y_0}{Z_+ \varepsilon_\mr{e} \sin_\mr{t} - \sin^2_\mr{t} \varepsilon_\mr{eh} E^y_0} \right) \right] \right\}^{-1}.
    \label{eqd:half_infinite_kerr}
\end{align}
Similarly to the 2D material, terms involving \eq{E^y_0} appear only in combination with \eq{\varepsilon_\mr{eh}} and give negligible corrections to the denominator. Therefore, we set \eq{E^y_0 = 0}, which gives \eq{\sin_\mr{t}^2 + \cos_\mr{t}^2 = 1} from Eqs.~(\ref{eqc:sint}, \ref{eqc:cost}), such that
\begin{equation}
    \Theta_\mr{s} = \frac{2Z_- \tan_\mr{t} \varepsilon_\mr{eh} E^x_0}{Z^2_+ \varepsilon_\mr{e} \left[1 - \frac{Z_- \cos_\mr{t}}{Z_+ \cos(\phi_\mr{i})} \right] \left[1 + \frac{Z_- \cos(\phi_\mr{i})}{Z_+ \cos_\mr{t}} \right]}.
\end{equation}
Equation~(12) of the main text is obtained by defining \eq{\cos_\mr{t} = C} and the refractive index \eq{n_+ = \sqrt{\varepsilon_\mr{e}}} while using \eq{\varepsilon_\mr{e} Z_+ = 1 / v_+} and the relation for \eq{\sin_\mr{t}} in Eq.~\eqref{eqc:sint}. 

\begin{figure}[h]
    \centering  \includegraphics[width=\linewidth]{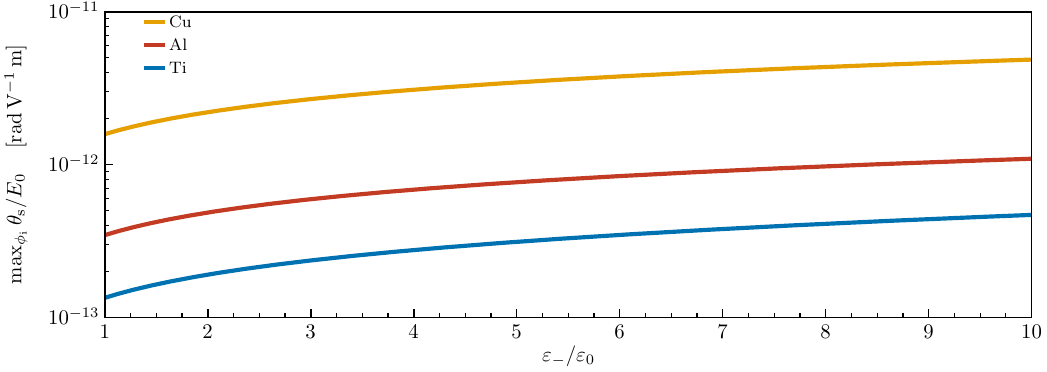}
    \caption{Maximum Kerr rotation angle per unit static electric field for copper (\eq{\mr{Cu}}), titanium (\eq{\mr{Ti}}), and aluminum (\eq{\mr{Al}}) modeled as semi-infinite materials, obtained by scanning over incidence angles \eq{\phi_\mr{i} \in (0, \pi / 2)}, vs. dielectric constant above the material \eq{(z < 0)}. The static electric field is \eq{\bm{E}_0 = \hat{\bm{x}} E_0} and the wavelength of the incident wave is \eq{\lambda = 800 \; \si{\nano\meter}}}.
    \label{figd:half_infinite_media}
\end{figure}

Figure~\ref{figd:half_infinite_media} shows that varying the dielectric constant above the material does not significantly influence the magnitude of the Kerr rotation angle.
\end{document}